\documentclass[12pt]{article}

\usepackage{scicite}

\usepackage{times}

\usepackage{graphicx}
\usepackage{amssymb}
\usepackage{multibib}
\usepackage{color}

\DeclareMathSymbol{\varOmega}{\mathord}{letters}{"0A}
\DeclareMathSymbol{\varPhi}{\mathord}{letters}{"08}
\DeclareMathSymbol{\varSigma}{\mathord}{letters}{"06}
\DeclareMathSymbol{\varPsi}{\mathord}{letters}{"09}
\DeclareMathSymbol{\varGamma}{\mathord}{letters}{"00}

\newcounter{lastnote}

\date{}

\begin{document} 

\baselineskip24pt

\noindent {\bf Title} \\
A pebble accretion model for the formation of the terrestrial planets in the
Solar System 

\noindent {\bf Authors} \\
Anders Johansen$^{1*,2}$, Thomas Ronnet$^{2}$, Martin Bizzarro$^{1}$, Martin
Schiller$^{1}$, Michiel Lambrechts$^{2}$, \AA ke Nordlund$^{3}$, \& Helmut
Lammer$^{4}$

\noindent {\bf Affiliations} \\
$^{1}$Center for Star and Planet Formation, GLOBE Institute, University of
Copenhagen, \O ster Voldgade 5-7, 1350 Copenhagen, Denmark \\
$^{2}$Lund Observatory, Department of Astronomy and Theoretical Physics, Lund
University, Box 43, 22100 Lund, Sweden\\
$^{3}$Niels Bohr Institute, University of Copenhagen, Juliane Maries Vej 30,
2100 Copenhagen, Denmark \\
$^{4}$Space Research Institute, Austrian Academy of Sciences, Schmiedlstr.~6,
8042 Graz, Austria

\noindent {\bf Abstract} \\
Pebbles of millimeter sizes are abundant in protoplanetary discs around young
stars. Chondrules inside primitive meteorites  -- formed by melting of dust
aggregate pebbles or in impacts between planetesimals -- have similar sizes. The
role of pebble accretion for terrestrial planet formation is nevertheless
unclear. Here we present a model where inwards-drifting pebbles feed the growth
of terrestrial planets. The masses and orbits of Venus, Earth, Theia (which
later collided with the Earth to form the Moon) and Mars are all consistent with
pebble accretion onto protoplanets that formed around Mars' orbit and migrated
to their final positions while growing. The isotopic compositions of Earth and
Mars are matched qualitatively by accretion of two generations of pebbles,
carrying distinct isotopic signatures. Finally, we show that the water and
carbon budget of Earth can be delivered by pebbles from the early generation
before the gas envelope became hot enough to vaporise volatiles.

\newpage

\noindent {\bf \large Introduction}

The long formation time-scale of gas giants and ice giants in the outer regions
of protoplanetary discs by traditional planetesimal accretion
\cite{Levison+etal2010,JohansenBitsch2019} instigated the development of the
pebble accretion theory in which the pebbles drifting through the protoplanetary
disc are accreted rapidly by the growing protoplanets
\cite{OrmelKlahr2010,LambrechtsJohansen2012}. While pebble accretion clearly
aids the formation of gas giants \cite{Johansen+etal2019}, the formation of
terrestrial planets has so far, with a few exceptions
\cite{Levison+etal2015,Johansen+etal2015,Popovas+etal2018,Popovas+etal2019},
mainly been explored in classical $N$-body simulations where terrestrial planets
grow by successive impacts between increasingly massive protoplanets
\cite{Raymond+etal2009}.

Observations of protoplanetary discs reveal that very young stars are orbited by
several hundred Earth masses of pebbles \cite{Tychoniec+etal2018}, embedded in a
gaseous disc of 100 times higher mass. This pebble population vanishes on a
characteristic time-scale of a few million years \cite{Ansdell+etal2016}, likely
due to a combination of radial drift and planetesimal formation outside gaps in
the gas caused by the gravity of the growing planets
\cite{Stammler+etal2019,Appelgren+etal2020,Eriksson+etal2020}. The orbital speed
of the gas in a protoplanetary disc is set by a balance between the gravity from
the central star and the outwards-pointing pressure force. The resulting
sub-Keplerian orbital motion of the gas acts as a headwind on the pebbles,
draining them of their orbital angular momentum and causing them to drift
radially towards the star \cite{Weidenschilling1977}. The inner regions of
protoplanetary discs, where terrestrial planets form, therefore witness a flow
of hundreds of Earth masses of pebbles throughout the life-time of the
protoplanetary disc.

The millimeter-sized chondrules found within primitive meteorites likely
represent pebbles that formed around the young Sun; the leading mechanism for
heating and melting these pebbles to form igneous chondrules is shock waves in
the solar protoplanetary disc \cite{DeschConnolly2002}. Alternatively,
chondrules may have formed from the molten debris of collisions between massive
protoplanets \cite{Johnson+etal2015}. Chondrules formed over the first 3 Myr of
the evolution of the protoplanetary disc
\cite{Connelly+etal2012,Bollard+etal2017} and dominated the mass budget when the
parent bodies of the ordinary chondrites meteorite class formed in the inner
regions of the solar protoplanetary disc $\sim$3 Myr after the formation of the
Sun. The large fraction of millimeter-sized chondrules in both ordinary
chondrites and carbonaceous chondrites is a strong indication that planetesimals
in the solar protoplanetary disc accreted in the gravitational collapse of
pebble clumps (possibly consisting of a mixture of individual chondrules and
larger aggregates of chondrules and dust) that could have formed through the
streaming instability. This hydrodynamical instability arises in the radial
drift flow of pebbles and causes the pebbles to form dense filaments that
collapse gravitationally into planetesimals with a characteristic size of 100 km
\cite{Johansen+etal2015,Simon+etal2016}. The largest planetesimals then continue
to accrete pebbles and grow to protoplanets with masses between the Moon and
Mars.

The radial flux of pebbles towards the star was recently demonstrated to
determine the outcome of planetary growth in the inner regions of the
protoplanetary disc \cite{Lambrechts+etal2019}: high pebble fluxes from the
outer protoplanetary disc lead to formation of migrating chains of super-Earths,
while any reduction in the pebble flux, e.g.\ by the emergence of giant planets
in the outer Solar System whose gravity on the gas disc acts as a pressure
barrier for pebble drift, strands the protoplanets at Mars-masses. This was
followed by a phase of giant impacts akin to classical terrestrial planet
formation.

In this paper we explore the possibility that the radial drift of pebbles
continues to drive the growth from protoplanets to masses comparable to Earth
and Venus by pebble accretion. The role of giant impacts for terrestrial planet
formation is here reduced to the Moon-forming impact between Earth and Theia, an
additional terrestrial planet that collided with Earth after an instability in
the system of primordial terrestrial planets that orbited our young Sun
\cite{QuarlesLissauer2015}. We show that a pebble accretion scenario for
terrestrial planet formation provides explanations for several properties of the
terrestrial planets in the Solar System, including (1) the masses and orbits of
Venus, Earth and Mars, (2) the isotopic composition of Earth and Mars and (3)
the delivery of carbon and water to Earth in amounts that are comparable to the
inferred reservoirs.

\begin{figure}[!h]
  \begin{center}
    \includegraphics[width=0.50\linewidth]{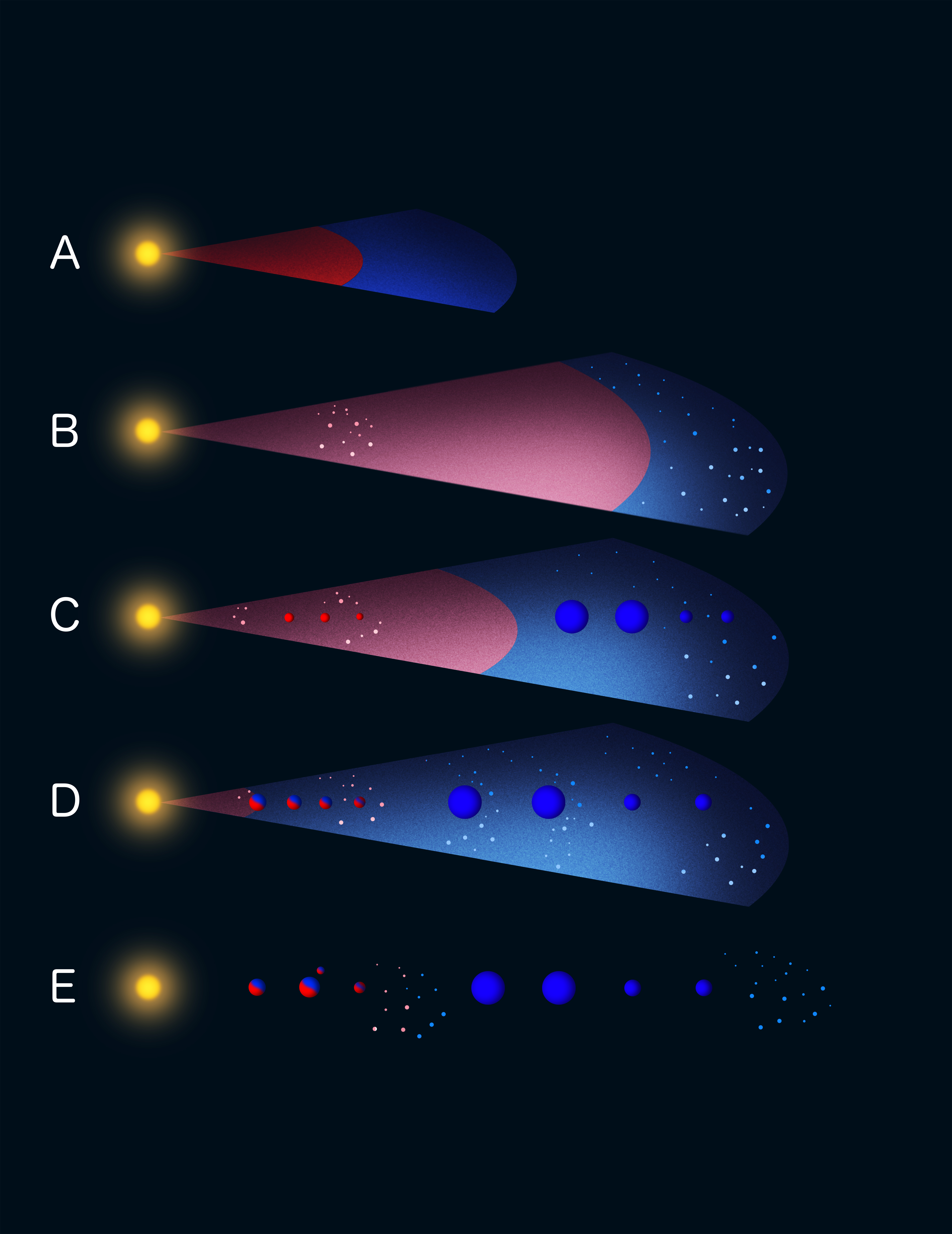}
  \end{center}
  \caption{{\bf Sketch of the physical processes involved in our pebble
  accretion model for the formation of terrestrial planets.} Stage (A): the
  protoplanetary disc is formed consisting of material with solar composition
  (blue), represented in the meteoric record by the CI meteorites. Thermal
  processing in the inner disc vaporizes presolar grains carrying isotopic
  anomalies. The remaining solids carry now a non-carbonaceous (NC) signature
  (red). In Stage (B) the disc expands outwards due to angular momentum
  transport from the inner to the outer disc, carrying the NC
  material along with the gas. Planetesimal belts form at the water ice line
  (red) and by pile ups of pebbles in the outer regions of the protoplanetary
  disc (blue); this outer planetesimal belt is envisioned here as the
  birth region of the giant planets \cite{Pirani+etal2019,Appelgren+etal2020}.
  In Stage (C) protoplanets representing Earth, Venus and Theia migrate out of
  the inner planetesimal belt. In the outer Solar System Jupiter, Saturn, Uranus
  and Neptune grow by pebble accretion and gas accretion. In Stage (D) the CI
  material has drifted past the terrestrial planet zone and the terrestrial
  planets shift their compositions more towards the CI meteorites.  The
  carbonaceous chondrites (CC) form outside of the orbits of Jupiter and Saturn.
  Finally, in stage (E) the protoplanetary disc clears and planetesimals of NC
  and CC composition are scattered into the asteroid belt.}
  \label{f:sketch}
\end{figure}

Our model, depicted as a sketch in Figure \ref{f:sketch}, provides important
constraints into the timing and flux of primitive outer disk material to the
inner Solar System. In particular, we conclude that Jupiter did not halt
inward mass transport of outer Solar System dust aggregates to the accretion
region of terrestrial planets, in contrast to models that invoke Jupiter as an
impermeable barrier that separated the inner and outer Solar System
\cite{Kruijer+etal2017}.

\noindent {\bf \large Results}

We develop and present here a model for terrestrial planet formation driven
dominantly by pebble accretion, but we include also self-consistently the
contribution from planetesimal accretion. We start by identifying the water ice
line as the most likely location for the formation of a first generation of
planetesimals that acted as seeds for pebble accretion in the solar
protoplanetary disc. As the stellar luminosity decreases with time, the ice line
moves interior to the region of terrestrial planet formation. This clearly has
implications for the delivery of water to terrestrial planets that form by
pebble accretion.

\noindent {\bf Location of the water ice line in the solar protoplanetary disc}

Planetesimal formation by the streaming instability requires that the ratio of
the surface density of pebbles relative to that of the gas is elevated above a
threshold metallicity, higher than the nominal solar value, in order to trigger
the formation of dense filaments that collapse to form planetesimals
\cite{Carrera+etal2015}. Such an increased pebble density has been demonstrated
to occur early in the evolution of the protoplanetary disc by pile up of dust
released by icy pebbles interior of the water ice line \cite{IdaGuillot2016} or
by the deposition of water vapor transported by diffusion to exterior of the ice
line \cite{SchoonenbergOrmel2017,DrazkowskaAlibert2017}. We therefore assume in
our model that an early generation of planetesimals formed at the water ice line
and that the planetesimals that grew to form Venus, Earth, Theia and Mars were
among them.

The orbital distance of the water ice line depends on the luminosity of
the young Sun as well as on the efficiency of viscous heating by the gas that is
accreted. We present here both a model where viscous heating is provided at all
distances from the star as well as a more realistic model where viscous heating
is only provided when the magnetorotational instability is active above $T=800$
K \cite{DeschTurner2015}. This ``dead zone'' model is described in details in
Materials and Methods.

\begin{figure}[!h]
  \includegraphics[width=0.50\linewidth]{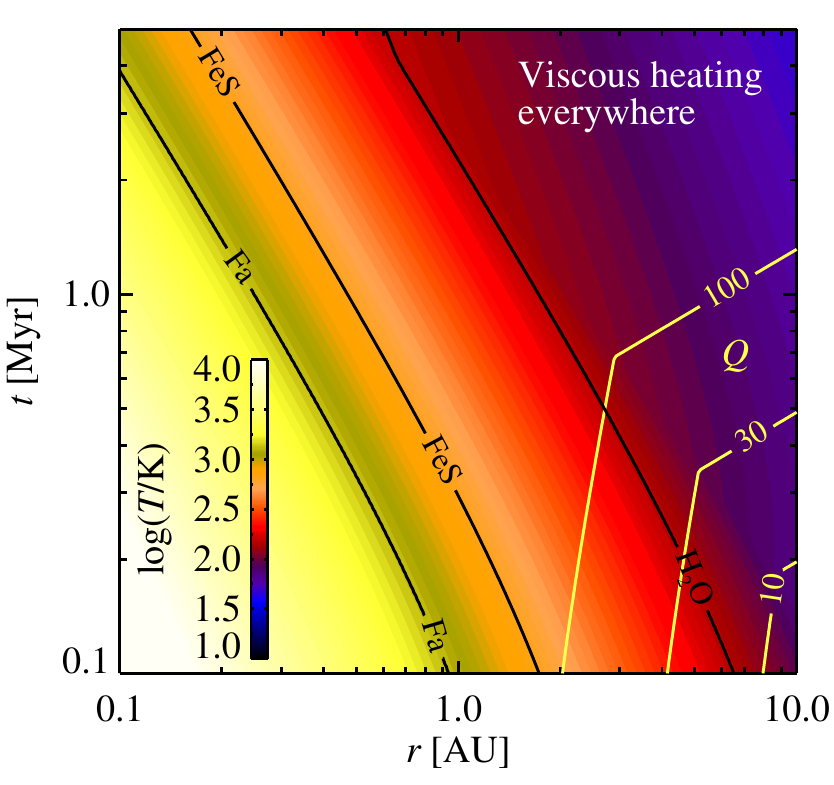}
  \includegraphics[width=0.50\linewidth]{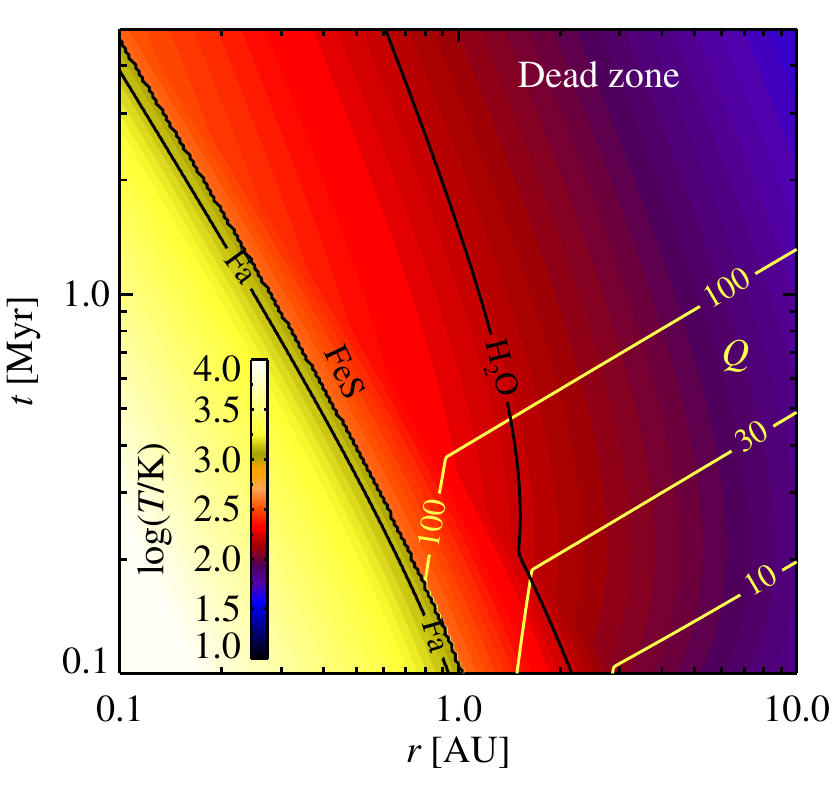}
  \caption{{\bf Temperature maps of the inner 10 AU of an evolving
  protoplanetary disc}. The left plot shows the temperature when viscous
  heating is applied everywhere in the protoplanetary disc. In the right plot
  we assume that viscous heat is only released when the magnetorotational
  instability is active above a temperature of 800 K -- the remaining disc is
  magnetically dead and hence only heated by the stellar irradiation. Three
  contour lines for the Toomre $Q$ parameter are indicated in yellow; values
  above $\approx$1 imply gravitational stability and hence validates our neglect
  of heating from gravitoturbulence.  We overplot the sublimation lines of water
  (H$_2$O) and the refractory minerals Troilite (FeS) and Fayalite
  (Fe$_2$SiO$_4$). In the more realistic case where viscous heating is provided
  only where the magnetorotational instability is active (right plot), the
  primordial water ice line sits in the region between 1.2 AU and 2.0 AU in the
  first million years of disc evolution.  This is the likely site for formation
  of the first generation of planetesimals in the inner Solar System.}
  \label{f:temperature_map}
\end{figure}

We assume that the protoplanetary disc column density evolves as a viscous
$\alpha$-disc from an initial accretion rate of $\dot{M}_{\rm
*0}=10^{-6}\,M_\odot\,{\rm yr^{-1}}$ at $t=0$ down to $\dot{M}_{\rm
*1}=10^{-9}\,M_\odot\,{\rm yr^{-1}}$ after 5 Myr. This decrease in the accretion
rate implies that the heating becomes more dominated by the stellar irradiation
with time. We show the calculated temperature of the solar protoplanetary disc
in Figure \ref{f:temperature_map}. Additional details of the structure of the
protoplanetary disc model with a dead zone are shown in Supplementary Figure
\ref{f:disc_structure}.

The ice line in the more realistic dead zone model sits initially in the
region between 1.2 and 2.0 AU, whereas the model with viscous heating everywhere
has a primordial ice line at 7 AU. As the luminosity of the Sun decreases with
the contraction of the protostar, the water ice line moves interior of 1 AU
after around 1.5--2 Myr in both models. Since the limiting case where viscous
heating is applied everywhere is not physically realistic, as demonstrated in
magnetohydrodynamical simulations \cite{Mori+etal2019}, we adopt the dead zone
model when integrating the growth tracks of planets. Overall we infer from the
dead zone model that the first planetesimals in the inner Solar Systems likely
formed in a region devoid of magnetically-driven turbulence between 1.2 and 2.0
AU.

\noindent {\bf Analytical growth tracks of terrestrial planets}

Protoplanets in the protoplanetary disc will grow at a rate $\dot{M}$ given by
the accretion of pebbles and planetesimals and migrate towards the star at a
rate $\dot{r}$. The combination of these two rates yields the growth track
equation ${\rm d} M/{\rm d}r = \dot{M}/\dot{r}$. This equation can be integrated
analytically \cite{LambrechtsJohansen2014,Johansen+etal2019} for the case of
growth by 2-D pebble accretion (relevant when the scale-height of the pebbles is
smaller than the pebble accretion radius) and standard, inwards type I planetary
migration to yield the analytical growth track expression
\begin{equation}
  M(r) = M_{\rm max} \left[ 1 - \left( \frac{r}{r_0} \right) ^{1-\zeta}
  \right]^{3/4} \, .
  \label{eq:Mt}
\end{equation}
We express the growth track here in the limit where $M \gg M_0$, where
$M_0$ is the starting mass. The gas temperature of the protoplanetary disc is
assumed to depend on the
distance from the Sun as $T \propto r^{-\zeta}$ and we take a constant value of
$\zeta=3/7$, valid for a protoplanetary disc that intercepts stellar irradiation
at a grazing angle \cite{Ida+etal2016}. This power law approximation
for the temperature holds in the dead zone of the protoplanetary disc where the
terrestrial planets grow and migrate. Equation (\ref{eq:Mt}) has two free
parameters: $r_0$ is the starting position of the protoplanet and $M_{\rm
max}$ is the mass of the protoplanet when it has migrated to $r=0$. We fit now
the growth track to Mars and Venus, assuming that their current masses reflect
their primordial masses when the solar protoplanetary disc dissipated. Earth
has suffered a giant impact that led to the formation of the Moon, so the
current mass of Earth does not represent its pre-impact mass. Matching the
masses and locations of Mars (index 1) and Venus (index 2) in equation
(\ref{eq:Mt}) allows us to divide out the $M_{\rm max}$ parameter to yield
\begin{equation}
  r_0^{\zeta-1} \left[ r_1^{1-\zeta} - (M_1/M_2)^{4/3} r_2^{1-\zeta} \right] =
  1 - (M_1/M_2)^{4/3} \, .
  \label{eq:r0}
\end{equation}
We see how the starting position, $r_0$, is dominantly set by the planet with
the smallest mass ($r_0 \rightarrow r_1$ for $M_1 \ll M_2$). For Mars and Venus
we obtain $r_0=1.59$ AU, slightly exterior of Mars' current orbit at $r=1.53$
AU. This location is broadly consistent with the likely location of the water
ice line during the first million years of protoplanetary disc evolution
when considering that the terrestrial planet zone was heated mainly by the
radiation from the central star. The maximum mass is then set by applying
equation (\ref{eq:Mt}) to Venus, giving $M_{\rm max}=1.74\,M_{\rm E}$.

\begin{figure}[!h]
  \includegraphics[width=0.47\linewidth]{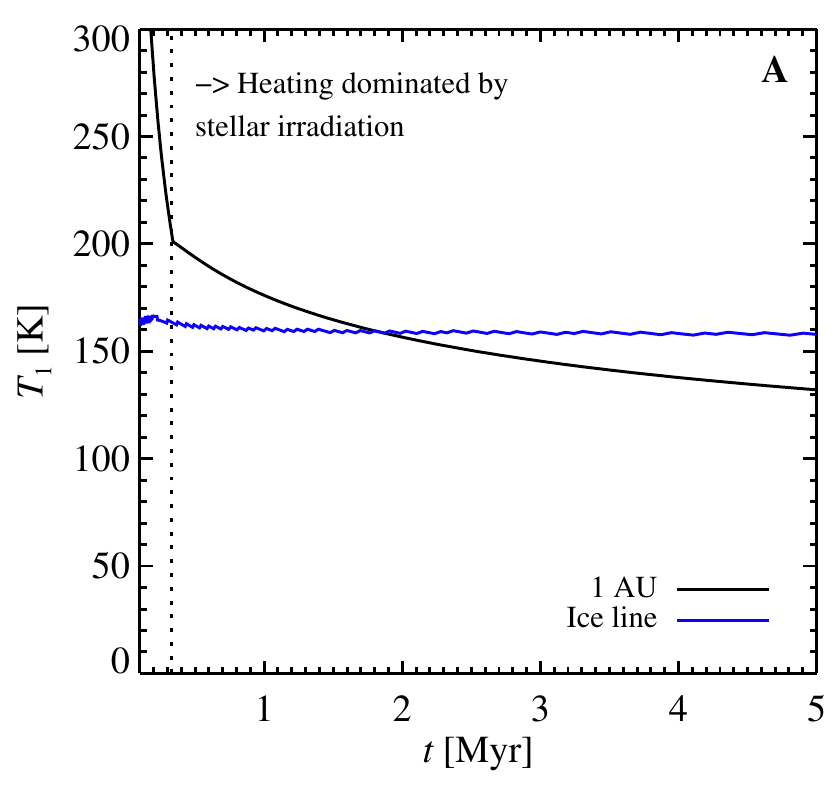}
  \includegraphics[width=0.50\linewidth]{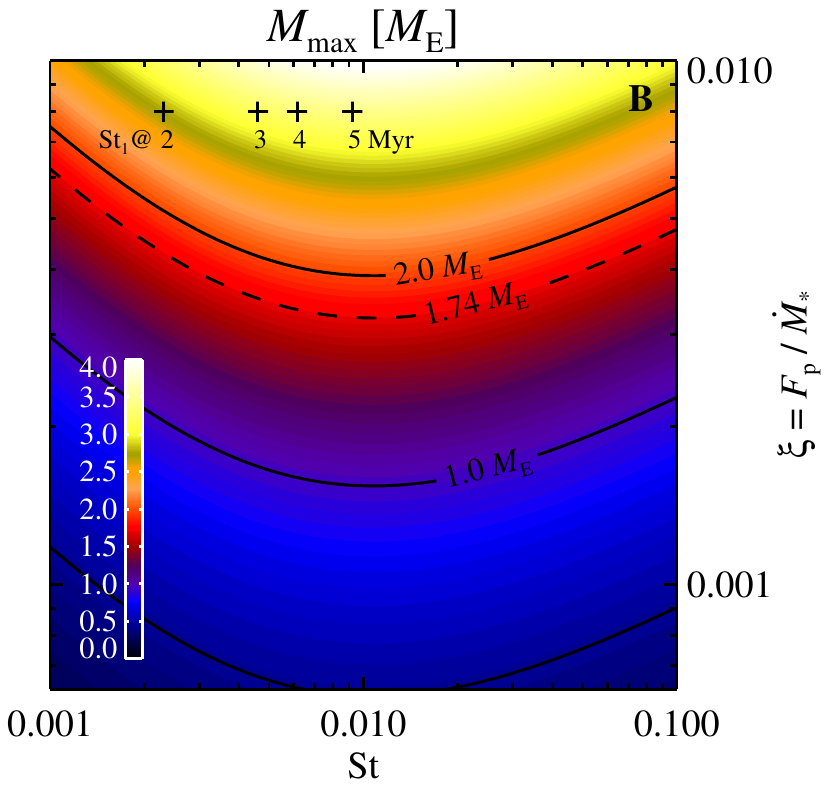}
  \caption{{\bf (A) The temperature at 1 AU distance from the star a
  function of time and the temperature of the water ice line.} The temperature
  becomes dominated by stellar irradiation already after 0.3 Myr of disc
  evolution; the temperature then continues to drop more slowly as the stellar
  luminosity falls with time. The temperature falls below water vapour
  saturation after approximately 2 Myr of disc evolution. {\bf (B) The
  maximum planetary mass for growth tracks starting at $r_0=1.6$ AU.} The mass
  is shown as a function of the pebble Stokes number, ${\rm St}$, and the ratio
  of the radial pebble flux rate through the protoplanetary disc relative to the
  gas flux rate, $\xi=F_{\rm p}/\dot{M}_\star$. We give the temperature a
  passive irradiation profile with fixed value of 140 K at 1 AU. The Stokes
  number at 1 AU for millimeter-sized pebbles are indicated at four different
  times. The maximum mass that leads to a good match for Venus' orbit and mass
  is indicated with a dashed line.  This is obtained for a range of Stokes
  numbers between 0.001 and 0.1, consistent with the evolution of the Stokes
  number of millimeter-sized pebbles from the early, dense protoplanetary disc
  to the late, dilute disc (pluses). The pebble-to-gas flux ratio lies in the
  range between 0.004 to 0.008; these values are similar to the solar ratio of
  refractory material relative to hydrogen/helium gas.}
  \label{f:Mmax_xi_St}
\end{figure}

Our successful fit of a single growth track to Mars and Venus is always
mathematically possible when the outer planet (Mars) is less massive than the
inner planet (Venus). How physically plausible this growth track is lies in the
value of the maximum mass $M_{\rm max}$.  This maximum mass, or migration mass,
depends on a number of parameters of the protoplanetary disc
\cite{Johansen+etal2019}. We focus here on the pebble Stokes number, ${\rm St}$,
a dimensionless number that characterizes the frictional stopping time of the
pebbles, and the ratio of the radial flux of pebbles relative to the flux of the
gas, $\xi=F_{\rm p}/\dot{M}_\star$, where $\dot{M}_\star$ is the gas accretion
rate onto the star. The maximum mass furthermore depends on starting
position $r_0$, which we fix here to 1.6 AU, and the temperature profile for
which we take a power law with index $\zeta=3/7$ and temperature $T_1=140$ K at
1 AU; we neglect any temporal dependence of the stellar luminosity. We
illustrate the dependence of the maximum mass on these parameters in Figure
\ref{f:Mmax_xi_St}. The maximum mass that yields the best fit to the orbit and
mass of Venus is obtained for a range of ${\rm St}$ from 0.001 to 0.1 and a
range of $\xi$ from 0.004 to 0.008. These ranges represent nominal values of the
Stokes number of millimeter-sized pebbles and radial pebble fluxes
\cite{Johansen+etal2019}.  From this we conclude that the masses and orbits of
Venus and Mars are consistent with growth by accretion of millimeter-sized
pebbles combined with standard type I planetary migration.

\noindent {\bf Numerical growth tracks of terrestrial planets}

After the successful application of the analytical growth tracks to Venus and
Mars, we now turn to numerical integration of the masses and orbits of
protoplanets undergoing pebble accretion, planetesimal accretion and inwards
type I migration, using the code developed and presented in
\cite{Johansen+etal2019} and \cite{JohansenBitsch2019}. The protoplanetary
disc model is described in Materials and Methods. Planetesimals with fixed
radii of 100 km are present from the beginning of the simulation as a Gaussian
belt of width 0.05 AU centred at 1.6 AU. The total planetesimal mass is
approximately 0.5 $M_{\rm E}$; this mass is broadly consistent with models
of early planetesimal formation at the water ice line in protoplanetary discs of
solar metallicity \cite{DrazkowskaAlibert2017}.

We start protoplanets representing Venus, Earth, Theia and Mars after $t=0.66$
Myr, $0.93$ Myr, $1.50$ Myr and $2.67$ Myr, respectively, after the formation of
the protoplanetary disc. The protoplanets are all given initial masses of $M_0 =
1.0 \times 10^{-3}$ $M_{\rm E}$. The starting times are chosen to be consistent
with the growth time from an initial mass function of planetesimals peaking at
$M \sim 10^{-7}$--$10^{-6}$ $M_{\rm E}$ to our starting mass of $M_0 =
10^{-3}\,M_{\rm E}$ by planetesimal collisions and pebble accretion
\cite{Johansen+etal2015}; the difference in growth time-scales for the
different bodies is essentially set by their different levels of eccentricity,
which controls the pebble accretion rate, until the bodies are circularized by
the transition to rapid pebble accretion at $M \sim 10^{-4}\,M_{\rm E}$
\cite{Johansen+etal2015}, approximately the mass of the largest asteroid Ceres.
We consider for simplicity pebble accretion only in the Hill regime, valid from
a factor 10 times lower than our starting mass \cite{LambrechtsJohansen2012}. We
ignore the gravitational interaction between the planets as they grow; in the
Supplementary Material we present the results of a full $N$-body simulation
showing that the single-planet growth tracks are reproduced when including the
gravity between the growing planets.

\begin{figure}[!h]
  \includegraphics{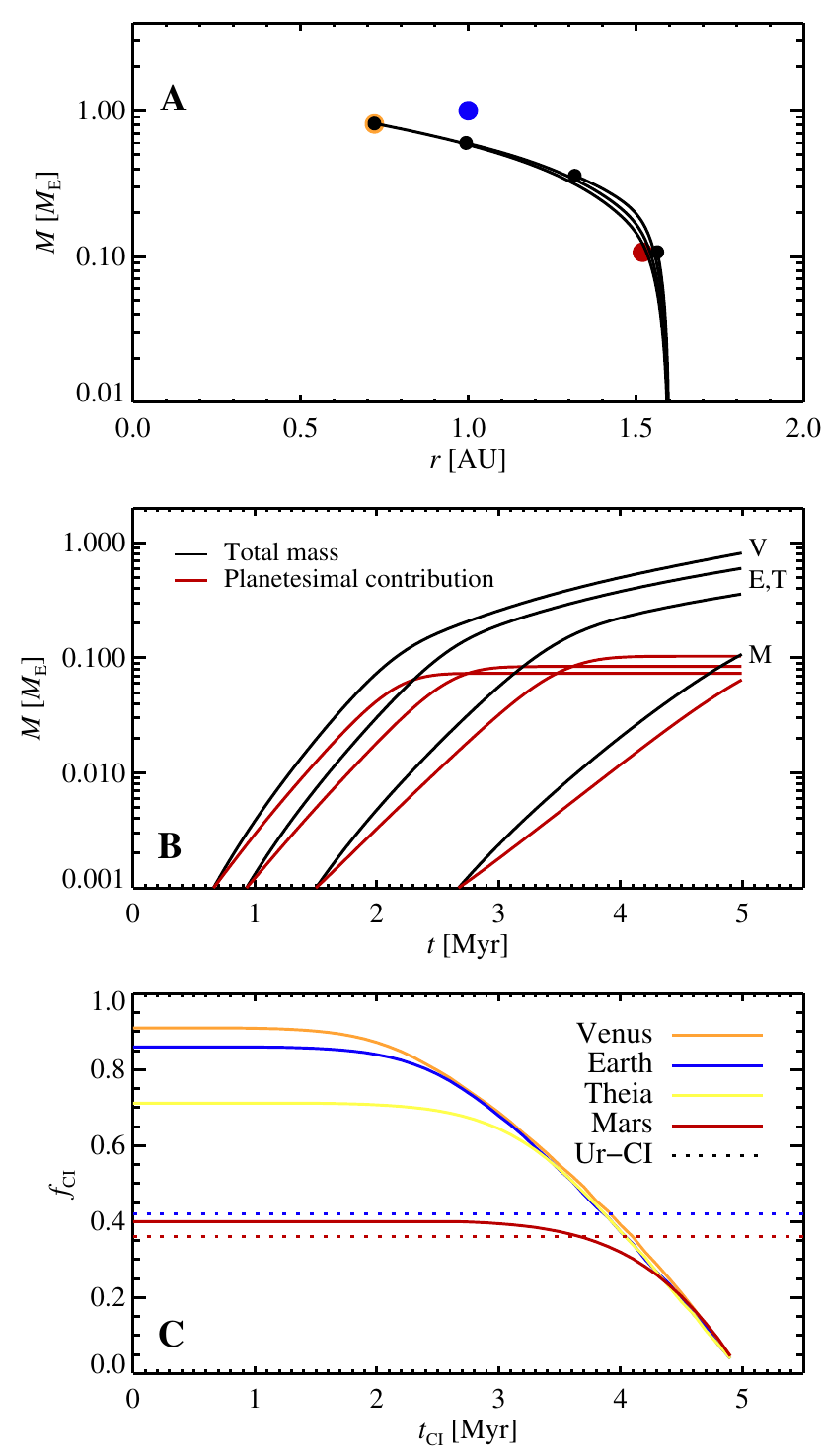}
  \caption{{\bf (A) Numerical growth tracks of protoplanets growing by pebble
  accretion and planetesimal accretion, starting at $r_0=1.6\,{\rm AU}$.} The
  parameters of the growth track were chosen with Venus and
  Mars as end members. Earth obtains a final mass of $0.6\, M_{\rm E}$, which
  we augment by creating Theia slightly later to reach a final mass of
  $0.4\,M_{\rm E}$. {\bf (B) Masses of the four planets as a function of time}.
  We show here also the contribution from planetesimal accretion in the birth
  belt. {\bf (C) The fraction of mass accreted from outer Solar System
  material}, as a function of the time when the drifting pebbles change to CI
  composition. We find agreement with the estimated CI contribution of 42\% to
  Earth and 36\% to Mars \cite{Schiller+etal2018} for a transition time from
  ureilite composition to CI composition in the range 3.5--4.0 Myr.}
  \label{f:growth_tracks_pebbles}
\end{figure}

In Figure \ref{f:growth_tracks_pebbles} we show growth tracks starting at $r_0 =
1.6\,{\rm AU}$.  The different starting times of the protoplanets make a
hierarchy of final masses and orbits. We confirm from the numerical integrations
that Venus and Mars belong to the same growth track, although Mars grows faster
than estimated because of accreting a significant mass fraction (approximately
60\%) in planetesimals from its birth belt. The other planets accrete a similar
contribution from planetesimals, but their total mass budget is dominated by
pebble accretion and hence they follow the analytical growth track better than
Mars. Earth obtains a final mass of $0.6\,M_{\rm E}$, but this mass is augmented
by our assumption that Theia was a fifth terrestrial planet with an original
orbit between Earth and Mars. We construct Theia to have a mass of $0.4\,M_{\rm
E}$.  This mass is broadly consistent with models for Moon formation that invoke
a very massive impactor to vaporize the Earth-Theia collision debris
\cite{Lock+etal2018}.

\noindent{\bf Isotopic composition of the planets}

The mass-independent isotopic compositions of various elements such as O, Ti, Cr
and Ca have been measured for Earth as well as for Mars (from martian
meteorites) and the major meteorite classes \cite{DauphasSchauble2016}, and
shows that variability exists between the various classes of meteorites and
terrestrial planets. This nucleosynthetic variability likely reflects either a
time-dependent supply of grains with different nucleosynthetic heritage to the
solar protoplanetary disc or, alternatively, a temperature-dependent destruction
of either pre-solar grains or grains condensed in the local interstellar medium
\cite{Trinquier+etal2009,Schiller+etal2018,Ek+etal2019}, resulting in a
composition dichotomy between the inner, hot regions and the outer, cold regions
of the solar protoplanetary disc.

Broadly speaking, the isotopic abundances of meteorites fall in two distinct
groups when plotting any two elements against each other. One group is
represented by the non-carbonaceous (NC) meteorites, including ordinary
chondrites, and the other group by the carbonaceous chondrites (CC). Earth sits
on a mixing line between the two components, together with the enstatite
chondrites \cite{Warren2011}, establishing that a compositional gradient exists
between the NC and CC reservoirs. Therefore, Earth could have formed either from
material akin to enstatite chondrites to yield the measured isotopic fingerprint
\cite{Dauphas2017} or from a combination of material akin to ordinary chondrites
mixed with material akin to carbonaceous chondrites
\cite{Warren2011,Schiller+etal2018}. The latter picture nevertheless opens the
question of how carbonaceous chondrites, which likely formed beyond the ice line
in the solar protoplanetary disc, could have been a major source of material for
the terrestrial planets.

Based on the mean isotopic abundances of the non-carbonaceous chondrite group on
the one hand and of the carbonaceous chondrite group on the other hand
\cite{Warren2011}, it has been estimated that Earth and Mars accreted between
30\%--43\% and 15\%--30\% mass from the latter group, respectively, when
considering simultaneously the isotopes $^{54}$Cr and $^{50}$Ti. Based on the
isotopes $^{54}$Cr and $^{17}$O, Earth and Mars should have accreted between
18\%--32\% and 2.6\%--18\% from the carbonaceous chondrite group, respectively.
Using instead the abundance of $^{48}$Ca in the ureilites and CI chondrites as
end members for the mixing, Earth requires a contribution of 42\% from pebbles
of CI composition and Mars 36\% \cite{Schiller+etal2018}. We used $^{48}$Ca
(with end members ureilites and CI chondrites) to match the composition of Earth
and Mars in Figure \ref{f:growth_tracks_pebbles}, but choosing other end members
from the non-carbonaceous and the carbonaceous groups would give qualitatively
similar results.

To calculate the contribution of material from the outer regions of the solar
protoplanetary disc to the terrestrial planets formed in our simulations, we
assume that the planetesimals in the birth belt as well as the pebbles (and
chondrules) that drift through the disc for the first $t_{\rm CI}$ (a variable
time) have the isotopic composition of the ureilites meteorites (an end member
of the non-carbonaceous meteorite group understood to represent the initial
inner disc composition \cite{Schiller+etal2018}), followed by an influx of
dust-aggregate pebbles of composition similar to CI chondrites for the last
$5.0\,{\rm Myr}-t_{\rm CI}$, before the protoplanetary disc finally dissipates
after 5 Myr.  The time $t_{\rm CI}$ can be chosen to fit best the measured
compositions of Earth and Mars. In panel (C) of Figure
\ref{f:growth_tracks_pebbles} we match \cite{Schiller+etal2018} in getting
approximately 42\% contribution of pebbles with CI composition to Earth for a
transition time $t_{\rm CI}=3.8\,{\rm Myr}$. In Supplementary Figure
\ref{f:alpha_disc_trajectories} we show that the CI material resided beyond 10
AU after $10^5$ yr of evolution of the protoplanetary disc, having a total gas
and dust mass of 0.027 $M_\odot$ out of the total disc mass of $0.044$ $M_\odot$
at this stage, and was subsequently pushed outwards by the viscous expansion of
the disc before falling back through the terrestrial planet region at $t_{\rm
CI}=3.8\,{\rm Myr}$. At this time the gas flux through the protoplanetary disc
is $\dot{M}_* \approx 2 \times 10^{-9}\,M_\odot\,{\rm yr}^{-1}$ and the radial
pebble flux is $F_{\rm p} \approx 2.4\,M_{\rm E}\,{\rm Myr}^{-1}$.  Venus and
Theia obtain very similar compositions to Earth, since all three planets accrete
from the same drifting material.

Mars would get a similar fraction as the other planets of non-carbonaceous
relative to carbonaceous material from pebble accretion; however Mars has an
isotopic composition that lies closer to the ordinary chondrites than Earth.
This could indicate that Mars terminated its growth earlier than Venus, Earth
and Theia and hence avoided the incorporation of pebbles of CI composition that
drifted in later \cite{Schiller+etal2018}. However, that raises the question of
how Mars' eccentricity and inclination could have been excited during the
gaseous disc phase. Instead, we assume here that all the four planets accrete
planetesimals within their birth belt. This contribution of material with inner
Solar System ureilite-like composition affects Mars the most, since it accretes
the largest fraction of its mass in the planetesimal birth belt. Hence, we can
qualitatively explain the compositional difference between Earth and Mars as a
natural consequence of Mars being less massive than Earth and thus having a
larger planetesimal accretion contribution to its mass. The compositions of
Earth and Mars are broadly reproduced by a pebble composition transition time
$t_{\rm CI}$ in the range between 3.5 and 4.0 Myr.

The isotopic signature of iron in the mantle of the Earth is very similar to
that of iron in CI chondrites \cite{Schiller+etal2020}. This agreement is
consistent with our model. The early-accreted iron from the first
generation of pebbles came from the NC reservoir and was hence largely reduced,
with iron present as metallic iron nuggets as in the ordinary chondrites. This
iron accreted together with ice until the envelope reached the ice sublimation
temperature at a mass of $0.02\,M_{\rm E}$ after approximately 2 Myr (see
sections on volatile delivery below). The iron
arriving together with ice would have oxidized in contact with liquid water in
the protoplanet, melted by the heating by $^{26}$Al, but this early oxidized
iron only constitutes a small fraction of the total oxidized iron found in the
mantle of the Earth today.  Material accreted later contained too little
$^{26}$Al to melt and hence the iron remained in solid form in a primitive
mantle overlying the earlier differentiated layers. The bulk of the metal of NC
composition separated from the silicates and entered the core as the protoplanet
finally melted by the accretion heat. The second generation of pebbles of CI
composition contained a large fraction of oxidised iron that remains in the
mantle today. This is in agreement with the abundances of siderophile
elements in the Earth's mantle which require increasingly oxidizing conditions
during silicate-metal separation with increasing mass of the growing Earth
\cite{Rubie+etal2015}. The oxidised iron in the Earth's mantle constitutes
around 16\% of the total iron on Earth, but the outer core likely contains a
significant fraction of oxidised iron as well \cite{Badro+etal2014}. Given our
approximately 40\% contribution from pebbles of CI composition to Earth, these
pebbles should be approximately 60\% oxidised in order to account for the iron
budget in the Earth's mantle. The iron in the CI chondrites is in fact nearly
100\% oxidised, but the oxidation of primordial FeS and metallic iron could have
occurred by liquid water flow in the parent body \cite{Endress+etal1996}.
The total mass of the oxidized iron the Earth's mantle implies furthermore
that the iron in the Earth's core consists of 25\% iron derived from CI-like
solids and 75\% iron derived from NC material.

Jupiter has been proposed to play an important role in the separation of
the NC and CC reservoirs in the solar protoplanetary disc
\cite{Kruijer+etal2017}.  However, hydrodynamical simulations show that the gap
formed by the combined action of Jupiter and Saturn in the protoplanetary disc
is permeable to small dust aggregates \cite{Haugbolle+etal2019}, particularly if
the gap edge develops turbulence e.g.\ through the Rossby wave instability
\cite{Lyra+etal2009}.  Larger pebbles will become trapped in the peak of the gas
pressure exterior to the gap -- but fragmenting pebble collisions continuously
create dust that can pass through the gap \cite{Drazkowska+etal2019}. Hence, in
our view Jupiter did not present an efficient barrier to the supply of dust
particles from the outer regions of the protoplanetary disc to the terrestrial
planet zone. Jupiter could nevertheless have acted to reduce the flux of
pebbles, consistent with the low pebble flux needed to explain the masses and
orbits of Venus, Earth, Theia and Mars in our model.

\noindent{\bf Free parameters in our model}

It is remarkable that it is possible to choose a starting position and four
starting times of the protoplanets and match the orbits, masses and compositions
of Venus, Earth, Theia and Mars by pebble accretion, including a contribution
from planetesimal accretion in the birth belt, using realistic parameters for
the pebble sizes and the radial pebble flux. Our model nevertheless has a number
of parameters that are to a varying degree free to choose. We briefly discuss
the choice of those here.

We distinguish between three classes of parameters: (a) free parameters that are
unconstrained from observations, (b) free parameters that are constrained from
observations and (c) fixed parameters that are constrained by observations and
set according to the most reasonable value. The truly free parameters in class
(a) are $t_0$ (the four starting times of the protoplanets when they have the
starting mass $M_0$), $r_0$ (the starting position of the protoplanets), $t_{\rm
CI}$ (time of infall of pebbles of CI composition) and $M_{\rm pla}$ (the total
planetesimal mass in the birth belt) and $\xi$ (the pebble mass flux
relative to the gas mass flux, which we fine tune around the solar composition
to
fit the growth tracks). The only free parameter in class (b) is $R_{\rm p}$
(the pebble size, we set it here to be 1 mm, similar to chondrule sizes and to
pebble sizes inferred from observations of protoplanetary discs). The fixed
parameters in class (c) are $\zeta=3/7$ (the temperature gradient in the
protoplanetary disc), $\dot{M}_{\rm *0}=10^{-6}\,M_\odot\,{\rm yr}^{-1}$ and
$\dot{M}_{\rm *1}=10^{-9}\,M_\odot\,{\rm yr}^{-1}$ (the initial and final
accretion rate of the protoplanetary disc), $t_{\rm disc}=5\,{\rm Myr}$ (the
life-time of the protoplanetary disc), $\alpha=10^{-2}$ (the turbulent viscosity
in the protoplanetary disc) and $\delta=10^{-4}$ (the diffusion coefficient of
the gas, which governs the degree to which the pebbles sediment to the
mid-plane). We chose reasonable values, based on observations of protoplanetary
discs, for all parameters in class (c) and did not attempt to vary them in order
to obtain better fits to the observed properties of the terrestrial planets.

The combined set of parameters describes a model that has as outcome the masses,
orbits and compositions of the planets Venus, Earth, Theia and Mars. Here Theia
is a special case, since its mass is only constrained to add up with the mass of
proto-Earth to obtain the current mass of the Earth. We therefore consider the
real prediction of our model to be $M_{\rm V}$ (the mass of Venus), $M_{\rm M}$
(the mass of Mars), $M_{\rm E}+M_{\rm T}$ (the total mass of proto-Earth and
Theia), $r_{\rm V}$ (the orbit of Venus), $r_{\rm M}$ (the orbit of Mars),
$f_{\rm CI,E}$ (the composition of Earth), $f_{\rm CI,T}$ (the composition of
Theia, must be equal to that of Earth) and $f_{\rm CI,M}$ (the composition of
Mars). These eight model predictions are constructed by essentially choosing
three starting times of the protoplanets (considering the starting times of
Earth and Theia to be degenerate), the starting position of the protoplanets,
the radial the pebble flux relative to the gas flux, the time of infall
of pebbles of CI composition and the total mass of planetesimals in the birth
belt. This gives a total of seven truly free parameters that are used to fit
the data. The fact that the number of parameters is smaller than the number of
successful predictions implies that the model has a genuine explanation power
for the properties of the terrestrial planets in the Solar System.

\noindent{\bf Monte Carlo populations}

We now explore the effect of varying the birth positions and times of the
protoplanets as well as the width of the planetesimal belt. We start 1,000
protoplanets with masses $M_0=10^{-3}\,M_{\rm E}$ and give them random starting
times between 0.5 Myr and 5.0 Myr. The positions are drawn from a Gaussian belt
centred at $r_0=1.6\,{\rm AU}$ and with a width of either $W=0.05\,{\rm AU}$ or
$W=0.5\,{\rm AU}$. The growth tracks of Figure \ref{f:growth_tracks_pebbles}
were integrated using the narrow belt. The wider belt has ten times more mass in
planetesimals than the narrow belt, so a total of approximately $5\,M_{\rm E}$.

\begin{figure}[!h]
  \includegraphics[width=0.45\linewidth]{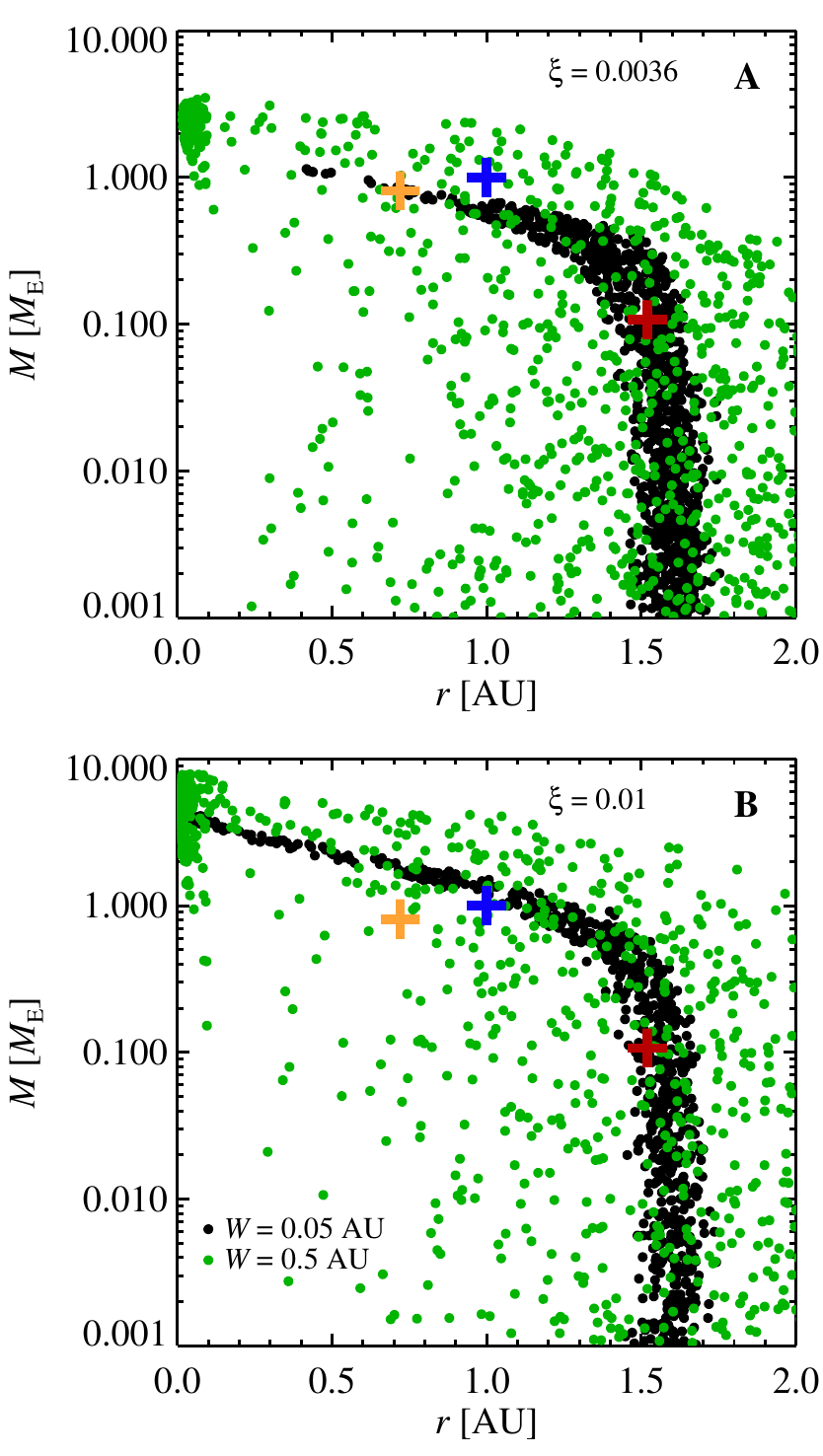}
  \caption{{\bf Monte Carlo sampling of 1,000 protoplanets starting with
  initial masses of $M_0=10^{-3}\,M_{\rm E}$ at random times between 0.5 Myr and
  5 Myr.} The black dots show results for a planetesimal belt of width
  $W=0.05\,{\rm AU}$ while the green dots show results of a width of
  $W=0.5\,{\rm AU}$. The protoplanets are started at random positions within the
  planetesimal belt.  (A) Results for a pebble-to-gas flux ratio of
  $\xi=0.0036$, as in Figure \ref{f:growth_tracks_pebbles}. The narrow belt case
  faithfully reproduces the characteristic masses and orbits of Venus and Mars.
  Considering instead a wider planetesimal belt leads to the formation of
  massive planets that migrate to the inner edge of the protoplanetary disc. (B)
  Results for a higher pebble-to-gas flux ratio of $\xi=0.01$. Venus is now well
  below the growth track, while planets in the Mars region grow to
  $\approx$$0.5\,M_{\rm E}$ before migration becomes significant. The high
  pebble flux allows super-Earths of up to $10\,M_{\rm E}$ to grow and migrate
  to the inner edge of the protoplanetary disc.}
  \label{f:growth_map}
\end{figure}

In Figure \ref{f:growth_map} we show the final positions and masses of the
considered protoplanets. We demonstrate results for two values of the pebble
flux: $\xi=0.0036$ as in Figure \ref{f:growth_tracks_pebbles} and a higher value
of $\xi=0.01$. The narrow belt case faithfully reproduces the characteristic
masses and orbits of Venus and Mars. Widening the planetesimal belt to 0.5 AU
increases the maximally reached planetary masses to between $2\,M_{\rm E}$ and
$5\,M_{\rm E}$ (super-Earths). The mass hierarchy with increasing planetary mass
with decreasing distance from the star is maintained as in the narrow belt case,
but the terrestrial planets no longer occupy characteristic positions in mass
versus semi-major axis.

The lower panel of Figure \ref{f:growth_map} shows the results when
increasing the pebble-to-gas flux ratio to $\xi=0.01$. This is closer to the
nominal flux in the absence of drift barriers, such as planets, further out in
the disc. The narrow belt case now fits Earth better, but Venus is well below
the characteristic growth track and planets at the location of Mars now reach
much higher masses before they start to migrate towards the star.  Super-Earths
of masses up to $10\,M_{\rm E}$ pile up at the inner edge of the protoplanetary
disc. The decisive role of the pebble flux in determining whether protoplanets
in the inner regions of a protoplanetary disc grow to super-Earths or ``only''
to Mars masses was already discovered in \cite{Lambrechts+etal2019}; here we
demonstrate that the combination of a low pebble flux and a narrow planetesimal
belt leads to the formation of analogues of the terrestrial planets in the Solar
System in terms of both their masses and orbits.

\noindent{\bf Sublimation of volatiles in the planetary envelope}

The accreted pebbles will carry molecules that are in solid form at the
temperature level of the surrounding protoplanetary disc. Since the water ice
line is interior of the accretion region of Earth and Venus during the main
accretion phase of the terrestrial planets, this implies that the pebbles will
be rich in volatiles. However, the growing planets attract gas envelopes that
are heated by the high pebble accretion luminosity and this heat will in turn
process the pebbles thermally while they are falling towards the surface of the
protoplanet. Our methods for calculating the temperature and density of the
envelope in hydrostatic and energy equilibrium are described in Materials and
Methods.

\begin{figure*}[!h]
  \includegraphics[width=0.49\linewidth]{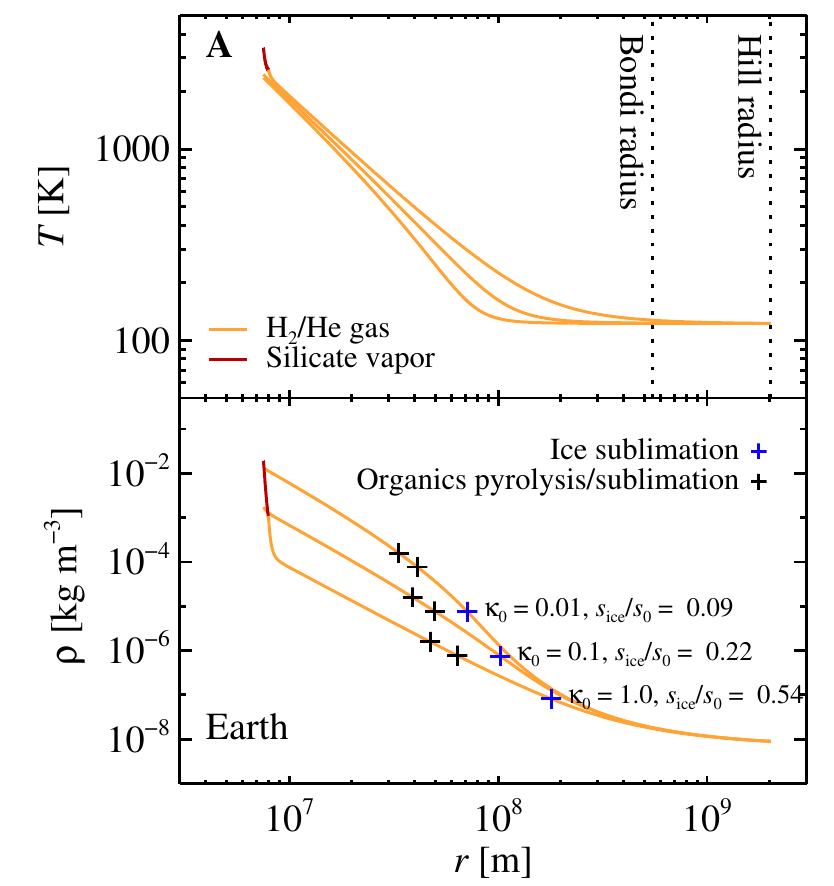}
  \includegraphics[width=0.49\linewidth]{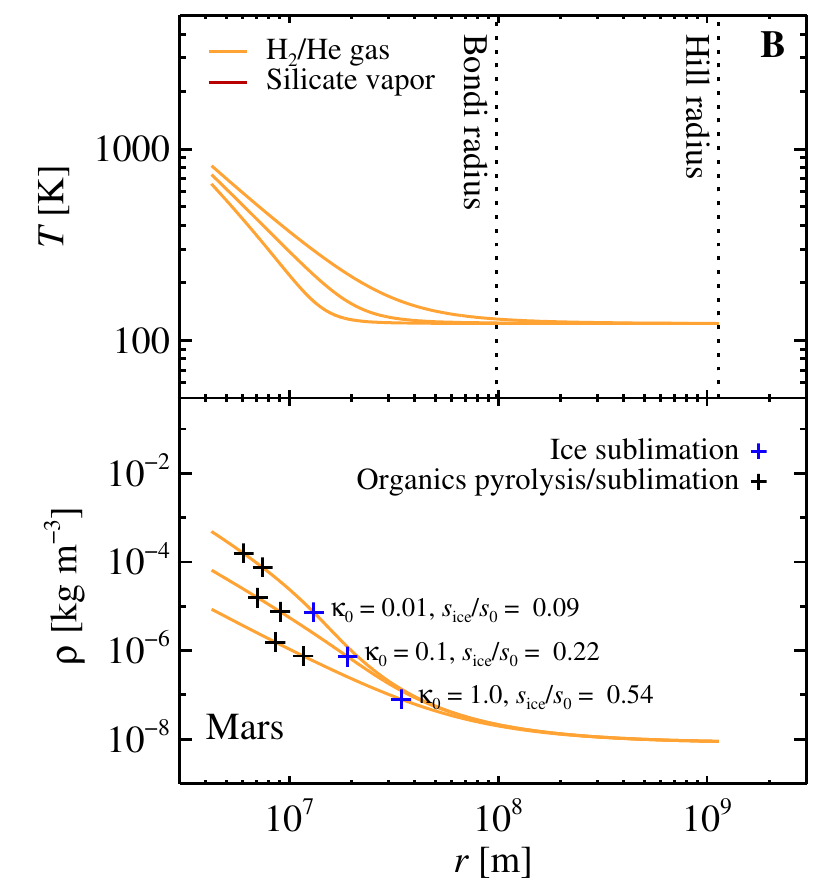}
  \caption{{\bf (A) Envelope structure of our Earth analogue and (B) our Mars
  analogue immediately prior to the dissipation of the protoplanetary disc at
  $t=5\,{\rm Myr}$}. The top panels show the gas temperature and the bottom
  panels the gas density.  Three different opacity levels are considered:
  $\kappa_0 = 0.01, 0.1, 1.0$ m$^2$ kg$^{-1}$. The temperature of the Earth's
  envelope directly over the magma ocean core reaches 2,000--3,000 K. The
  saturated vapor pressure of silicates (assumed here to be Forsterite)
  dominates over the ambient pressure only in a tiny region above the surface
  that reaches temperatures above approximately 2,600 K. We mark the relative
  entropy level at the water ice line. For the high-opacity case the relative
  entropy approaches 50\% of the disc entropy; flows from the protoplanetary
  disc easily reach this entropy level and cleanse the isothermal region of
  water vapor and ice particles.  Mars remains colder than 1,000 K throughout
  the envelope, and avoids melting, but the water ice line lies at a similar
  entropy level to Earth. The envelopes of both Earth and Mars reach
  temperatures in the range 325--425 K slightly below the water ice lines; here
  organic molecules undergo pyrolysis and sublimation to form volatile gas
  species (CH$_4$, CO and CO$_2$). These species can not recondense and will
  freely diffuse back to the protoplanetary disc.}
  \label{f:envelope_structure_Earth}
\end{figure*}

In Figure \ref{f:envelope_structure_Earth} we show the temperature and density
of the envelopes of our Earth and our Mars analogues at $t=5\,{\rm Myr}$ as a
function of the distance from the center of protoplanet, for three different
values of the opacity $\kappa_0$ (see Materials and Methods). The density of the
hydrogen/helium envelope rises exponentially inwards from the Bondi radius
(defined as $R_{\rm B} = G M / c_{\rm s}^2$ where $c_{\rm s}$ is the sound speed
of the gas in the protoplanetary disc). For Earth, the envelope temperature lies
above the 2,600 K of silicate sublimation close to the planetary core. This high
surface temperature will lead to extensive melting of the protoplanet and to the
formation of a magma ocean of at least 1,000 km in depth \cite{Lammer+etal2020}.
Our Mars analogue does not reach silicate melting temperatures at the surface,
but trapped heat released by $^{26}$Al within the protoplanet could heat the
interior above the melting temperature.

We indicate in Figure \ref{f:envelope_structure_Earth} the location of the water
ice line with blue pluses labelled with the entropy, relative to the entropy
level of the surrounding protoplanetary disc, at the depth of the water ice
line.  Hydrodynamical simulations have demonstrated that flows from the
protoplanetary disc reach down to a relative entropy level of $s/s_0 \approx
0.2$ \cite{KurokawaTanigawa2018}. Hence, for the high-opacity cases, with
$\kappa_0 = 0.1\,{\rm m^2\,kg^{-1}}$ and $1.0\,{\rm m^2\,kg^{-1}}$, the water
vapor that is released by sublimating the ice layers from the pebbles will be
recycled back to the protoplanetary disc.  Such a high opacity is easily reached
for a dust-to-gas ratio around the solar value or higher.

The recycling flows will first transport water vapor back across the ice line,
where the vapor nucleates on tiny dust particles that carry the dominant
surface area. Whether these ice-covered particles eventually reach the
protoplanetary disc depends on the time-scale to coagulate to large enough sizes
to sediment back through the gas flow. The coagulation time-scale of the ice
particles can be calculated from the approximate expression
\begin{equation}
  \tau_{\rm c} \sim \frac{a_{\rm ice} \rho_\bullet}{(\Delta v) \rho_{\rm ice}}
  \, .
\end{equation}
Here $a_{\rm ice}$ is the radius of the particles, $\rho_\bullet$ is their
material density, $\Delta v$ is the particle collision speed and $\rho_{\rm
ice}$ is the spatial density of the nucleated ice particles. We then assume that
the collision speed is dominated by brownian motion. That yields
\begin{equation}
  \tau_{\rm c} \sim \frac{a_{\rm ice} \rho_\bullet}{\sqrt{k_{\rm B} T/m}
  \rho_{\rm g}} \epsilon^{-1} \sim 4.2 \,{\rm yr}\, \left( \frac{a_{\rm
  ice}}{\mu{\rm m}} \right)^{5/2} \left( \frac{T}{170\,{\rm K}} \right)^{-1/2}
  \left( \frac{\rho_{\rm g}}{10^{-6}\,{\rm kg\,m^{-3}}} \right)^{-1} \left(
  \frac{\epsilon}{0.01} \right)^{-1} \, .
\end{equation}
We defined here $\epsilon=\rho_{\rm ice}/\rho_{\rm g}$ as the ratio of the
density of ice particles to the gas. The coagulation time-scale should now be
compared to the time-scale for the recycling flow to transport the particles
from the ice line at $R_{\rm ice}$ out to the Hill radius with the
characteristic speed $v_{\rm rec} \sim \varOmega R_{\rm ice}$. That gives
\begin{equation}
  \tau_{\rm rec} \sim (R_{\rm H}/R_{\rm ice})\varOmega^{-1} \sim 10
  \varOmega^{-1} \sim 2.0\,{\rm yr} \, .
\end{equation}
This time-scale is comparable to the coagulation time-scale around the ice line,
but the coagulation time-scale increases steeply in the decreasing gas density
exterior of the ice line. Hence, the freshly nucleated ice particles do not
coagulate appreciably during their transport from the ice line to the Hill
radius with the recycling flows.

\noindent {\bf Water delivery by pebble accretion}

\begin{figure}[!h]
  \includegraphics[width=0.5\linewidth]{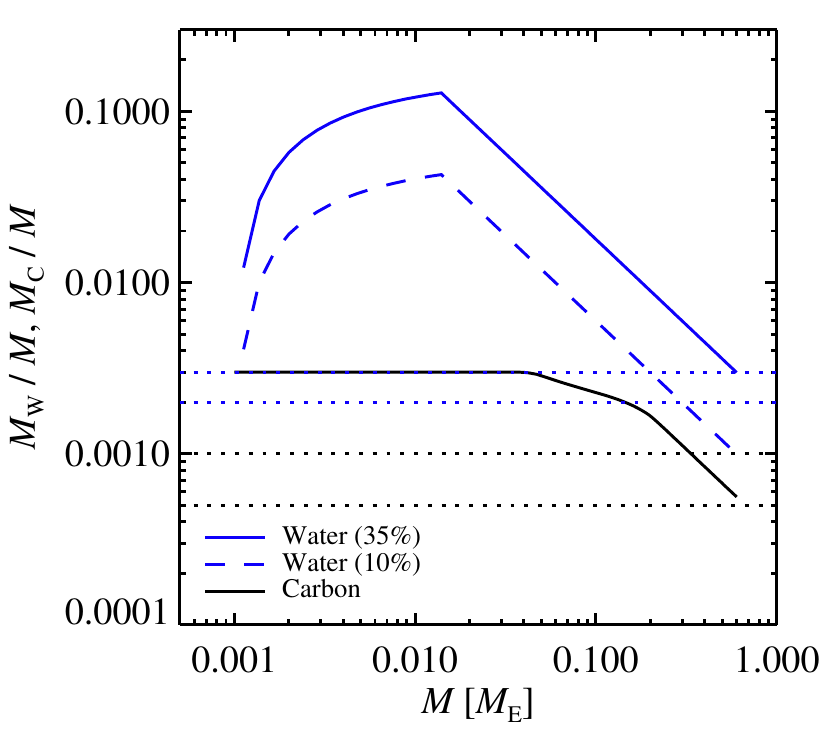}
  \caption{{\bf Fraction of water and carbon that survives the passage through
  the planetary envelope.} The fraction is shown as a function of the mass of
  the growing Earth. The results are shown for two different values of the ice
  fraction of the pebbles. The estimated water and carbon fractions of Earth are
  indicated with dotted lines. Water is delivered by icy pebbles until the
  temperature in the envelope reaches ice sublimation (assumed here to be at 160
  K). The water fraction falls steeply after the protoplanet reaches
  $0.02\,M_{\rm E}$. Carbon is vaporised in two steps -- organics vaporise in
  the temperature range 325-425 K and carbon dust burns at 1,100 K.
  Planetesimals in the birth belt as well as the early-accreted pebbles define
  an initial carbon fraction of 3,000 ppm. The infall of pebbles of CI
  composition (5\% carbon, assumed here to occur at $t_{\rm CI}=3.8\,{\rm Myr}$)
  when Earth reaches a mass of $0.3\,M_{\rm E}$ coincides with the increased
  heating of the planetary envelope, so that the overall carbon fraction
  actually decreases with increasing mass. The final value lands around 600 ppm,
  similar to estimates for the bulk Earth composition including a potential
  carbon reservoir in the core \cite{Marty2012,Chen+etal2014}.}
  \label{f:carbon_water_Earth}
\end{figure}

Figure \ref{f:carbon_water_Earth} shows the water fraction of our growing Earth
analogue, assuming that water can only be delivered to the protoplanet in the
form of ice accreted before the envelope reaches the temperature to sublimate
water ice. We assume that the accreted planetesimals are dry (since they formed
early enough to dry out due to the heat from $^{26}$Al, see reference
\cite{Lichtenberg+etal2019}) and that the accreted pebbles carry either
the nominal 35\% ice by mass, which assumes that some oxygen is bound in
the volatile CO molecule \cite{Bitsch+etal2019}, or a much dryer 10\%. The
water fraction peaks at a protoplanet mass just below $0.02\,M_{\rm E}$
(approximately the mass of the Moon) and then falls steeply for higher masses as
water ice is sublimated in the hot envelope. The final water fraction of our
Earth analogue is 3,000 ppm for the nominal ice fraction (35\%) and 1,000 ppm
for the low ice fraction (10\%).  The nominal case lands thus at the high end of
the 2,000-3,000 ppm estimated for Earth \cite{Marty2012}. Theia with a final
mass of $0.4\,M_{\rm E}$ acquires 4,000 ppm of water and would increase the
total water fraction of Earth to 3,500 ppm after the giant impact, but this
value could be reduced by loss of volatiles in energetic the impact.

All the water is delivered as ice in the early stages of protoplanet growth when
the protoplanet is still solid. The protoplanet will melt later by accretion
heat and radioactive decay and hence the magma ocean inherits a significant
fraction of water. Parts of this water will later be released from the magma
during crystallisation \cite{Elkins-Tanton2011} and likely mixed with water
formed from hydrogen ingassed from the protoplanetary disc \cite{Wu+etal2018}.
Any water that is outgassed during the growth of the protoplanet will be
retained as a high-density primary atmosphere below the hydrogen/helium
envelope, protected against loss to diffusive convection by the gradient in the
mean molecular weight between the atmosphere and the envelope. Our model thus
shows that water could have been delivered to Earth from ``pebble snow''
accreted during the early growth stages when the protoplanet was still solid and
the envelope cold enough for water ice to survive the passage to the surface.

\noindent {\bf Carbon delivery by pebble accretion}

In Figure \ref{f:envelope_structure_Earth} we also indicate the temperatures
where organic compounds are vaporised. We take the range between 325 K and 425
K for the vaporisation of 90\% of the carbon by pyrolysis and sublimation
\cite{GailTrieloff2017}. Pyrolysis and sublimation of organics converts these
molecules to very volatile carbon-bearing molecules such as CH$_4$, CO$_2$ and
CO. These species will not condense in the envelope and can diffuse freely to
the recycling zone. We assume that protoplanetary disc flows penetrate into the
Hill sphere with the characteristic speed $\varOmega R_{\rm B}$ at the Bondi
radius $R_{\rm B}$. This results in a mass loss rate of carbon-bearing molecules
as
\begin{equation}
  \dot{M}_{\rm C} = 4 \pi R_{\rm B}^2 \rho_{\rm C} \varOmega R_{\rm B} \, .
\end{equation}
We now equate this expression with the mass accretion rate of carbon, $f_{\rm C}
\dot{M}$, and isolate the density of carbon-bearing molecules, $\rho_{\rm C}$,
at the Bondi radius to be
\begin{eqnarray}
  \rho_{\rm C} &=& \frac{f_{\rm C} \dot{M}}{4 \pi R_{\rm B}^3 \varOmega} = 1.1
  \times 10^{-11}\,{\rm kg\,m^{-3}} \left( \frac{f_{\rm C}}{0.05} \right) \left(
  \frac{\dot{M}}{0.2\,M_{\rm E}\,{\rm Myr^{-1}}} \right) \left(
  \frac{M}{0.5\,M_{\rm E}} \right)^{-3} \nonumber \\
  & & \times \left( \frac{c_{\rm s}}{7.0 \times
  10^2\,{\rm m\,s^{-1}}} \right)^6 \left( \frac{r}{\rm AU} \right)^{3/2} \, .
\end{eqnarray}
This density is much lower than the gas density in the protoplanetary disc,
$\rho_{\rm g} \sim 10^{-8}\,{\rm kg\,m^{-3}}$, and hence the equilibrium ratio
of volatile carbon molecules relative to the gas is much lower than the ambient
density of CO, the main carbon-bearing molecule in inner regions of the
protoplanetary disc. The equilibrium mixing ratio at the Bondi radius,
$\rho_{\rm C}/\rho_{\rm g}$, will be maintained in the entire envelope by
turbulent diffusion with coefficient $D$. This diffusion equilibrium is
maintained after a characteristic time-scale of
\begin{equation}
  \tau = \frac{R_{\rm B}^2}{D} \, .
\end{equation}
The characteristic value of the gas mass in the envelope of terrestrial planets
that grow by pebble accretion is $M_{\rm g} \sim 10^{-4}\,M_{\rm E}$. The
time-scale to match this gas mass with carbon-bearing molecules released in the
organics sublimation/pyrolysis region is
\begin{equation}
  \tau_{\rm C} = \frac{M_{\rm g}}{f_{\rm C} \dot{M}} = 1 \times 10^4 \,{\rm
  yr}\, \left( \frac{M_{\rm g}}{10^{-4}\,M_{\rm E}} \right) \left( \frac{f_{\rm
  C}}{0.05} \right)^{-1} \left( \frac{\dot{M}}{0.2\,M_{\rm E}\,{\rm Myr}^{-1}}
  \right) \, .
\end{equation}
The turbulent diffusion time-scale must be shorter than this value to avoid
building up a layer in the envelope dominated by carbon-bearing molecules, which
would lead to an increase in the mean molecular weight of the gas and hence
potentially protect the carbon-bearing molecules from turbulent diffusion.
Setting the turbulent diffusion coefficient $D = v_{\rm turb} \times R_{\rm B}$
we get a turbulent speed of
\begin{equation}
  v_{\rm turb} = \frac{R_{\rm B}}{\tau} = 1.3 \times 10^{-3}\,{\rm m\,s^{-1}}
  \left( \frac{M}{0.5\,M_{\rm E}} \right) \left( \frac{c_{\rm s}}{7.0 \times
  10^2\,{\rm m\,s^{-1}}} \right)^{-2}  \left( \frac{\tau}{1 \times 10^4\,{\rm
  yr}} \right)^{-1} \, .
\end{equation}
This is an extremely modest speed compared to the sound speed and hence we
conclude that the carbon-bearing species released by vaporisation of organics
will become transported to the recycling flows of the protoplanetary disc before
they can build up a substantial mass fraction in the envelope.

Only approximately 10\% of the carbon provided by the pebbles will therefore
make it below the region where organics are vaporised. The remaining pure carbon
dust burns at 1,100 K \cite{GailTrieloff2017}. This temperature is reached
approximately 10,000 km above the surface of our Earth analogue. Assuming that
the carbon is only retained if it makes it to the protoplanet's surface, we
estimate the total carbon delivery to our Earth analogue as follows. We assume
that the carbon contents of the early generation of pebbles is similar to
ordinary chondrites (approximately 0.3\%) and similar to CI chondrites for the
late generation (approximately 5\%). The result is shown in Figure
\ref{f:carbon_water_Earth}. The fraction of carbon is initially constant at
3,000 ppm (set by the carbon contents of ordinary chondrites), but falls
steadily after carbon vaporisation starts in the envelope. Our Earth analogue
ends up with 600 ppm of carbon.

The total amount of carbon in our planet is relatively poorly constrained.
Earth's bulk carbon may reside in the core, adding up to a total of $(3-6)
\times 10^{21}\,{\rm kg}$ or 500-1,000 ppm \cite{Marty2012}. This
is within range of our estimate for the carbon delivery integrated over the
growth of the Earth. Our Mars analogue does not reach high enough temperatures
to burn carbon dust in its envelope and therefore obtains a carbon fraction as
high as 2,000 ppm. The more massive planets (Venus, Earth and Theia) accrete a
substantial amount of carbon in their earlier growth phases, before entering a
phase of burning carbon dust in the envelope after growing beyond $0.2\,M_{\rm
E}$. This carbon-free accretion phase lowers the resulting carbon fraction by
approximately a factor of three.

\noindent {\bf \large Discussion}

We built our model for terrestrial planet formation by pebble accretion around
fundamental physical processes that have been explored within the context of
modern planet formation theory. The water ice line has been demonstrated as a
likely site for early planetesimal formation; the location of the primordial
water ice line in the 1--2 AU region is a direct consequence of the elevated
luminosity of the young Sun, compared to its current value, and fits broadly
with the location of Mars. Taking Mars as a planet that underwent only modest
growth by pebble accretion as one end member and Venus that experienced
substantial growth and migration as the other end member, we demonstrated that
the masses and orbits of these two planets are consistent with growth by
accretion of millimeter-sized pebbles in a protoplanetary disc with a relatively
low radial flux of solid material. The radial pebble flux needed to match the
masses and orbits of the terrestrial planets corresponds to approximately
$10\,M_{\rm E}\,{\rm Myr}^{-1}$ at a stellar accretion rate of
$10^{-8}\,M_{\odot}\,{\rm yr}^{-1}$ (after 1 Myr of evolution in our model) and
$1\,M_{\rm E}\,{\rm Myr}^{-1}$ when the protoplanetary disc dissipates at an
accretion rate of $10^{-9}\,M_{\odot}\,{\rm yr}^{-1}$ after 5 Myr. These low
pebble fluxes correspond well to the estimated $1-10\,M_{\rm E}$ of solids left
in protoplanetary discs at the 3--5 Myr evolution stages
\cite{Ansdell+etal2016}.  Although such evolved protoplanetary discs may not be
able to form gas-giant planets any longer, they could presently be in their main
phase of forming terrestrial planets by pebble accretion.

The mass, orbit and composition of Mercury do not fit within this picture.
Instead, we propose that Mercury formed by accretion of metallic pebbles and
planetesimals outside of the sublimation front of Fayalite (Fe$_2$SiO$_4$),
which we demonstrate in Figure \ref{f:temperature_map} crosses Mercury's current
orbit after 0.5 Myr of disc evolution. This process is similar to how icy
pebbles trigger the streaming instability exterior of the water ice line
\cite{SchoonenbergOrmel2017}. Growth by pebble and planetesimal accretion is
very rapid so close to the Sun, so the current mass of Mercury is reached within
a few hundred thousand years. Mercury then ceased to accrete as the inner
regions of the solar protoplanetary disc underwent depletion by disc winds
\cite{Ogihara+etal2018}, stranding Mercury with a dominant metallic core and a
mantle whose silicates were reduced by accretion of pebbles rich in sulphur
released at the Troilite (FeS) sublimation front \cite{Jacquet+etal2015}.
Mercury could thus be the oldest of the terrestrial planets, having compiled its
bulk mass within a million years of the formation of the Sun.

Within our model framework, Earth reaches 60\% of its current mass at the
dissipation of the protoplanetary disc after 5 Myr. This mass of the proto-Earth
agrees well with constraints from matching the $^{36}$Ar/$^{38}$Ar,
$^{20}$Ne/$^{22}$Ne and $^{36}$Ar/$^{22}$Ne ratios by drag from the hydrodynamic
escape of the primordial hydrogen/helium envelope \cite{Lammer+etal2020}. We
augment the mass of Earth with the introduction of an additional planet Theia
(40\% Earth mass) between the orbits of Earth and Mars, to later collide with
the Earth to form our Moon. The radial drift of pebbles naturally provides Earth
and Theia with very similar compositions, even when the isotopic signature of
the drifting pebbles changes with time. Hence, the Earth and the Moon inherit
the indistinguishable isotopic compositions of their parent planets.

Models of protoplanetary discs show that the water ice line inevitably passes
interior to 1 AU as the stellar luminosity decreases with time
\cite{Ida+etal2016}. This implies that protoplanets growing in the terrestrial
planet zone will accrete a significant mass fraction of ice. However, ice is
sublimated in the hot gas envelope once the protoplanet mass reaches
$0.02\,M_{\rm E}$ and the water then escapes back to the protoplanetary disc
with the recycling flows that penetrate into the Hill sphere.  Volatile delivery
filtered through a hot envelope gives a good match to the water contents of
Earth. The early-accreted water ice becomes incorporated into the magma ocean
after the Earth melts by the accretion heat and the water will later degas as a
dense vapor atmosphere upon crystallisation of the Earth
\cite{Elkins-Tanton2011}, forming the first surface water masses on our planet.

The water on our Earth analogue is accreted as ice in the first 2 Myr of the
evolution of the protoplanetary disc, before the planetary envelope becomes hot
enough to sublimate the ice from the accreted pebbles. Since water is delivered
with pebbles from the early generation to Venus, Theia and Mars as well, we
predict that the primordial water delivered to Mars should have the same D/H
ratio as water on Earth. The early generation of pebbles in our model is
represented by the non-carbonaceous meteorite group, which includes the ordinary
chondrites. Water in the ordinary chondrites is heavier than Earth's water
\cite{Alexander+etal2012}, although a recent study found that enstatite
chondrites have a D/H ratio similar to water in the Earth's mantle
\cite{Piani+etal2020}. The D/H ratio of the ordinary chondrites could have been
increased significantly by oxidation of iron on the parent body, a process that
releases light hydrogen \cite{Sutton+etal2017}. The relatively dry parent bodies
of the ordinary chondrites could have formed at the warm side of the water ice
line after 3 Myr, when the ice line was interior of the current orbit of Venus.
Some of these would later have been dynamically implanted into the asteroid belt
by scattering, now seen as the S-type asteroids in the inner regions of the
asteroid belt \cite{RaymondIzidoro2017}.

Carbon, in the form of organics and carbon dust, is more refractory than water
and can reach the planetary surface for envelope temperatures up to 1,100 K
(where carbon dust burns). Carbon is thus delivered to the growing protoplanets
at masses up to $0.2\,M_{\rm E}$, after which the envelope becomes hot enough to
vaporize all the incoming carbon-bearing molecules. Our Earth analogue accretes
carbon at up to a time of 3.5 Myr. As with the water, the isotopic composition
of the carbon will therefore reflect material in the inner regions of the solar
protoplanetary disc. The overall carbon delivery filtered through the planetary
envelopes gives a good match to the estimated carbon reservoirs on Earth,
including a carbon component residing in the core.

Hence both water and carbon -- essential ingredients for life -- may have been
delivered to Earth in the form of ``pebble snow'' in the early phases of the
growth of our planet. This implies that volatile delivery need not be a
stochastic effect of a few giant impacts \cite{Marty2012} but may instead follow
predictable patterns that can be calculated from the evolution of the
protoplanetary disc and the radial drift of pebbles. If millimeter-sized pebbles
are the main providers of volatiles to planets in the inner regions of
protoplanetary discs, then we predict that super-Earths (more massive
counterparts of terrestrial planets found frequently around solar-like stars)
would have the same overall volatile budgets as terrestrial planets, being
unable to accrete any additional volatiles due to the intense heat in their gas
envelopes. The accretion of dense atmospheres of hydrogen and helium poses
an additional challenge to the habitability of super-Earths.

We believe that our pebble-driven model of terrestrial planet formation has
several advantages over the classical models that employ collisions between
planetary embryos as the main driver of planetary growth. While the latter
successfully produce planetary systems similar to the Solar System's terrestrial
planets in terms of masses and orbits, incorporating outer Solar System material
into the growing planets by giant impacts is very challenging. Simulations of
water-delivery to terrestrial planets employ collisions with wet protoplanets
originating beyond 2.5 AU -- the assumed ice line based on the dichotomy between
the S-type asteroids, associated with the ordinary chondrites, that dominate the
inner regions of the asteroid belt and the C-type asteroids, associated with the
carbonaceous chondrites, that dominate the outer regions. This position of the
ice line is in clear conflict with models of irradiated protoplanetary discs
showing that the ice line was likely situated interior of 1 AU for most of the
life-time of the disc. Notwithstanding this discrepancy, the classical models
provide the terrestrial planets with a water fraction between 0.1\% and 1\%.
This fraction can then be considered the maximum contribution from volatile-rich
outer Solar System material in the classical model -- and hence it stands in
contrast to the 40\% contribution needed to explain the isotopic composition of
the Earth based on $^{48}$Ca \cite{Schiller+etal2018} and particularly to the
match of the iron isotopic composition of the Earth's mantle with the CI
chondrites \cite{Schiller+etal2020}. Models invoking the inwards migration of
Jupiter through the asteroid belt to explain the small mass of Mars yield
similarly low fractions of outer Solar System volatile-rich material in the
terrestrial planets \cite{Rubie+etal2015}.

As an alternative view, the Earth could have formed from a local reservoir of
material that carried the exact same elemental and isotopic composition as
measured on Earth. The enstatite chondrites have been proposed as candidate
material matching Earth's isotopic composition, although the iron isotopes stand
out as an exception \cite{Schiller+etal2020}. Such an approach nevertheless
necessitates that all the terrestrial planets formed from local reservoirs of
distinct composition, perhaps a remnant of the condensation sequence in the
initially hot solar protoplanetary disc \cite{Bond+etal2010} although this view
conflicts with the volatile-rich mantle of Mercury \cite{Peplowski+etal2011}.
However, explaining the isotopic differences between the terrestrial planets by
local reservoirs highlights the problem of the similarity of the Earth and the
Moon, as Theia would thus be expected to have a distinct composition different
from that of Earth. In the pebble-driven model, Earth and Theia naturally get
very similar compositions, since they accrete from the same radial flux of
drifting pebbles.

\noindent {\bf \large Materials and Methods}

\noindent {\bf Protoplanetary disc model}

We consider a protoplanetary disc model where the temperature is set by
both irradiation from the host star and by viscous heating from the
magnetorotational instability. The latter is active when the temperature is
above 800 K \cite{DeschTurner2015}. We therefore increase the turbulent
viscosity that controls viscous heating from a background level of $\alpha_{\rm
v}=10^{-4}$ to $\alpha_{\rm v}=10^{-2}$, relevant for fully active turbulence
caused by the magnetorotational instability, over the temperature range from 800
K to 1,000 K.  We calculate the temperature structure iteratively by solving the
transcendental equation $T = T(\alpha_{\rm v})$ \cite{Ida+etal2016} where
$\alpha_{\rm v}$ is a function of $T$. We follow the database of I.~Baraffe
\cite{Baraffe+etal2015} to calculate the luminosity evolution of the young Sun.
We fit the Sun's luminosity evolution as $L(t) = 2 L_\odot (t/{\rm Myr})^{-0.7}$
for the times 0.5-5 Myr, where $L_\odot$ is the luminosity of the modern Sun. We
then multiply this value by a factor $\exp\{-1/[t/(0.15\,{\rm Myr})]\}$ to
represent the increase in stellar luminosity (including emission from accretion
and from deuterium burning, but not including the luminosity of the disc itself)
in the early stages of the stellar collapse phase \cite{YoungEvans2005}; this
effectively gives a peak in the protostellar luminosity of $L \approx 3 L_\odot$
at $t=0.2-0.3$ Myr, followed by a power law decline ending at $L \approx 0.5
L_\odot$ at $t=5$ Myr. The protoplanetary disc evolves via the
$\alpha$-viscosity prescription and the pebble mass flux through the disc is set
to a fixed fraction $\xi$ of the radial gas flux; this assumption of a constant
flux ratio is justified by growth bottlenecks in the viscously expanding outer
regions of the protoplanetary disc \cite{Johansen+etal2019}. The temporal
evolution of the structure of the protoplanetary disc is shown in Figure
\ref{f:disc_structure}.

The pebble column density $\varSigma_{\rm p}$ follows from setting the pebble
flux, $F_{\rm p} = 2 \pi r v_r \varSigma_{\rm p}$, equal to $\xi \dot{M}_*$
where $\dot{M}_\star$ is the gas accretion rate through the disc.  Here $v_r$ is
the radial drift speed of the pebbles, which is calculated self-consistently
from the temperature and pressure gradient of the gas in the protoplanetary
disc. The pebbles have a fixed size of 1 millimeter; this choice of pebble size
is motivated by the characteristic size of chondrules, particle growth to the
bouncing barrier \cite{Zsom+etal2010} and observations of protoplanetary discs
\cite{Carrasco-Gonzalez+etal2019}. We have checked that the
fragmentation-limited particle size is larger than 1 mm at 1 AU for all the
evolution stages of the adopted protoplanetary disc, for a turbulence strength
of $\delta=10^{-4}$ and a critical fragmentation speed of 1 m/s
\cite{Birnstiel+etal2012}. The pebble column density for $\xi=0.0036$ is
indicated in Figure \ref{f:disc_structure}.  The pebbles have Stokes numbers in
the range 0.001-0.01 (see Figure \ref{f:Mmax_xi_St}); for these small sizes the
drift speed is low enough that $\xi$ approximately represents the ratio of the
column density of pebbles relative to the gas. The pebble accretion radius is
calculated based on the expressions provided in \cite{JohansenLambrechts2017}
and are integrated in either 3-D (when the pebble scale-height is larger than
the accretion radius) or 2-D (when the pebble scale-height is smaller than the
accretion radius). To keep the model simple, we do not decrease the pebble
accretion rate when the ice layers are sublimated in the envelope at protoplanet
masses above $0.02\,M_{\rm E}$.

\noindent {\bf Calculating the equilibrium structure of the envelope}

We solve the structure of the planetary envelope by integrating the standard
equations of hydrostatic balance and outwards luminosity transport from the Hill
sphere down to the center of the protoplanet. The boundary conditions at the
Hill sphere are the temperature and density of the protoplanetary disc. We set
the temperature gradient to the minimum of the radiative temperature gradient
and the convective temperature gradient. Our consideration of the equilibrium
structure of the envelope ignores the time needed to heat the material in the
envelope to its equilibrium value. The accretion luminosity is generally high
enough to validate this approach. The radiative temperature gradient is defined
by the opacity; we assume here that the opacity follows a power law with the
temperature, $\kappa=\kappa_0 (T/100\,{\rm K})^{0.5}$ with $\kappa_0=0.01$,
$0.1$, or $1.0$ ${\rm m^2\,kg^{-1}}$ \cite{BellLin1994}. The mean molecular
weight of the gas is set to a weighted average of a solar mixture of
hydrogen/helium and silicate vapor at its saturated vapor pressure.

The total luminosity generated by the accretion, the latent heat of sublimation
of silicates and the decay of $^{26}$Al is
\begin{equation}
  L = \frac{G M \dot{M}}{R_{\rm sil}} - Q_{\rm sil} \dot{M} + L_{26} \, .
\end{equation}
Here $R_{\rm sil}$ is the either the location of the silicate sublimation front
(which is calculated self-consistently by equating the saturated vapor pressure
of silicates with the ambient pressure) or the surface of the planet (if the
temperature does not reach silicate sublimation). The latent heat of
sublimation of silicate rock ($Q_{\rm sil}$) reduces the accretion energy when
the temperature reaches silicate sublimation in the envelope. We
nevertheless ignore the latent heat of water and silicates in the calculations,
since silicate sublimation only happens at planetary masses approaching
$1\,M_{\rm E}$ while water sublimation is followed immediately by nucleation
of the outwards transport water vapour. The luminosity generated by the decay of
$^{26}$Al is set to
\begin{equation}
  L_{26} = r_{26} E_{26} M \, .
\end{equation}
Here $r_{26} = 1.3 \times 10^5\,{\rm kg^{-1}\,s^{-1}} \exp(-t/\tau)$ is the
decay rate of $^{26}$Al per total silicate mass \cite{Larsen+etal2016}, $t$ is
the time, $\tau=1.03\,{\rm Myr}$ is the decay constant, $E_{26} = 3.12\,{\rm
MeV}$ is the energy released per decay \cite{Castillo-Rogez+etal2009} and $M$ is
the mass of the silicate core. We assume that all the heat from $^{26}$Al is
released in the core, since the total mass of silicate vapor in the envelope is
much lower than the core mass for the planetary masses that we consider.

\noindent{\bf \large Supplementary Materials}\\
www.sciencemag.org\\
Materials and Methods\\
Supplementary Text\\

\nocite{scilastnote}
\bibliography{bibliography}
\bibliographystyle{ScienceAdvances.bst}

\noindent{\bf \large Acknowledgments}

\noindent{\bf General} \\
This project started after discussions held at the Europlanet and International
Space Science Institute Workshop ``Reading Terrestrial Planet Evolution in
Isotopes and Element Measurements'' held in Bern in October 2018.

\noindent{\bf Funding} \\
A.J.\ acknowledges funding from the European Research Foundation (ERC
Consolidator Grant 724687-PLANETESYS), the Knut and Alice Wallenberg Foundation
(Wallenberg Academy Fellow Grant 2017.0287) and the Swedish Research Council
(Project Grant 2018-04867). M.B.\ acknowledges funding from the Carlsberg
Foundation (CF18\_1105) and the European Research Council (ERC Advanced Grant
833275-DEEPTIME). M.S.\ acknowledges funding from the Villum Foundation (grant
number \#00025333). H.L.\ acknowledges funding from the Austrian Science Fund
(NFN project S11601-N16 and the related NFN subproject S11607-N16).

\noindent{\bf Author contributions} \\
The original idea for the project resulted from discussions between all of the
co-authors, with equal contributions. A.J.\ performed the computer simulations
in the main paper, analysed the data and wrote the manuscript. T.R.\ developed
and performed the $N$-body simulations presented in the Supplementary Material.
All authors contributed to finalising the original manuscript written by A.J.

\noindent{\bf Competing interests} \\
The authors declare that they have no competing interests

\noindent{\bf Data and materials availability} \\
All data needed to evaluate the conclusions in the paper are present in the
paper and/or the Supplementary Materials. Additional data related to this paper
may be requested from the authors.

\clearpage

\noindent {\bf \large Figures and Tables} \\


\newpage

\newpage

\newpage

\newpage

\newpage

\newpage

\newpage

\setcounter{figure}{0}
\renewcommand{\thefigure}{S\arabic{figure}}

\section*{Supplementary Materials}

\section{Supplementary Text}

{\bf N-body simulations of terrestrial planet formation by pebble accretion}

To verify the calculations of single growth tracks presented in the main text we
present here $N$-body simulations of terrestrial planet formation by pebble
accretion. The $N$-body simulations are performed with the {\it rebound} code,
available at \texttt{http://github.com/hannorein/}. We use the hybrid MERCURIUS
integrator and a time-step of $5 \times 10^{-3}\,{\rm yr}/(2 \pi)$. The pebble
accretion efficiency is calculated based on \cite{LiuOrmel2018} and
\cite{OrmelLiu 2018}. For the damping of eccentricity, inclination and
semi-major axis by gravitational torques from the gas disc, we follow
\cite{CresswellNelson2006} with $k_{\rm mig}=1$.

The gas surface density is identical to \cite{Lambrechts+etal2019} with
$\varSigma_{\rm g} = 610\,{\rm g\,cm^{-2}}(r/{\rm AU})^{-1/2}\exp(-t/\tau)$,
aspect ratio $h = 0.024 (r/{\rm AU})^{2/7}$ and the pebble flux $F_{\rm peb} =
40\,M_{\rm E}\,{\rm Myr}^{-1}\exp(-t/\tau)$. Here $\tau=1.5\,{\rm Myr}$ is the
accretion time-scale of the protoplanetary disc. The pebbles are given Stokes
numbers ${\rm St} = 10^{-3}\exp(t/\tau)$. This exponential model is different
from the $\alpha$-disc model presented in the main text, but we recover similar
masses and orbits of the terrestrial planets.

The protoplanets are started at the positions $r_0=(1.585,1.6,1.63,1.65)\,{\rm
AU}$, to avoid overlapping orbits from the beginning. The starting times are
$t_0=(0.77,0.95,1.15,2.78)\,{\rm Myr}$ and the starting masses are
$M_0=(10^{-3},10^{-3},10^{-3},10^{-2})\,M_{\rm E}$. We ignore in the $N$-body
simulations the accretion of planetesimals from the birth belt. Therefore we
start Mars with a ten times higher mass than we use for the other protoplanets,
in order to mimic the substantial contribution of planetesimal accretion to
Mars' mass that we find in the main text.

\begin{figure*}[!h]
  \begin{center}
    \includegraphics[width=0.90\linewidth]{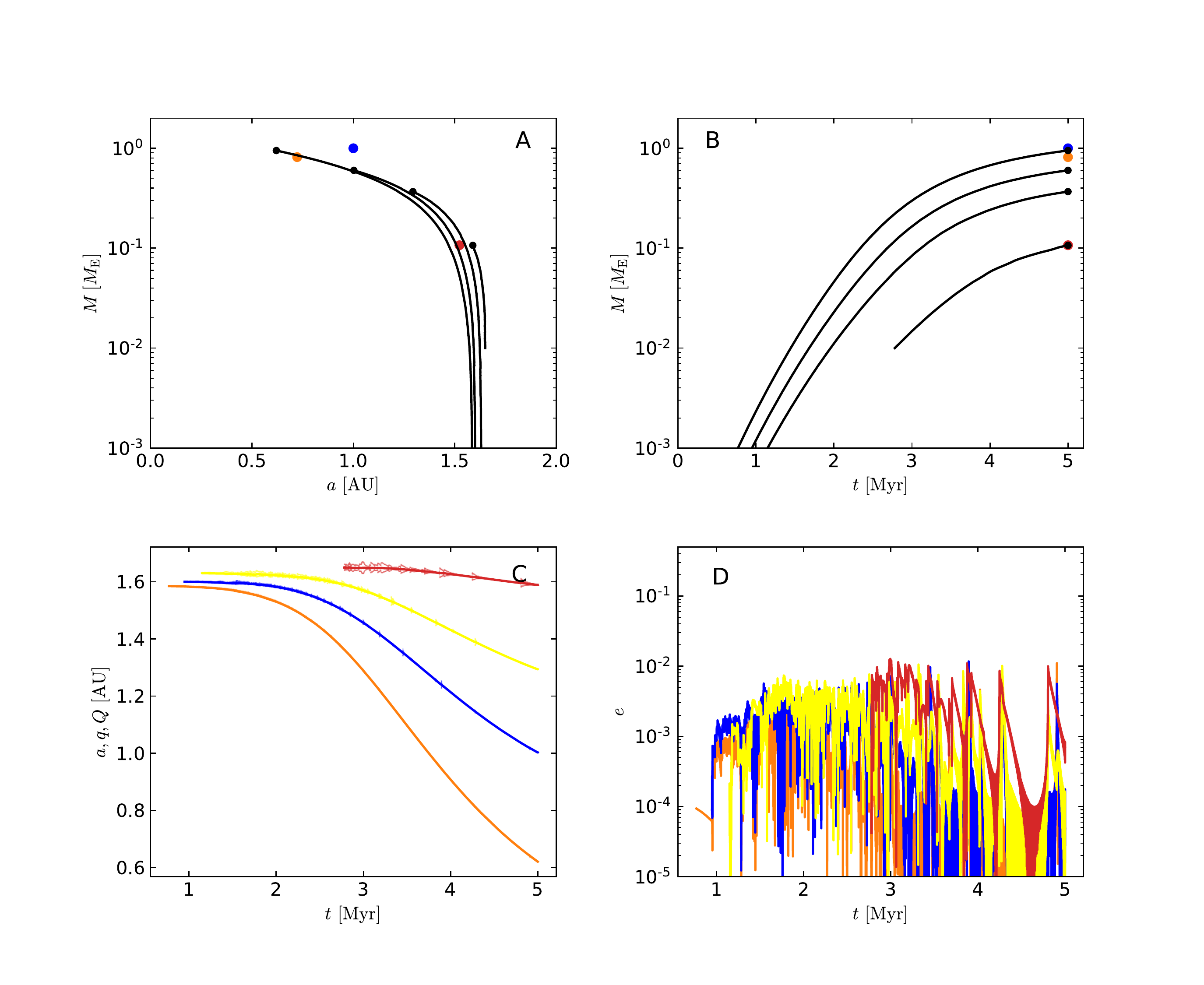}
  \end{center}
  \caption{{\bf Results of $N$-body simulations of terrestrial planet formation
  from protoplanets that start around $r_0=1.6\,{\rm AU}$ in the protoplanetary
  disc.} Panel (A) shows the growth tracks.  The growth tracks are similar to
  the single-planet growth tracks presented in the main paper. Panel (B) shows
  the planetary masses versus time, with a very similar shape again to the
  results in the main paper. Panel (C) shows the semi-major axes as a function
  of time. Here the interactions between the protoplanets is clear in the thin
  line that marks perihelion and aphelion distance of the orbit. In Panel (D) we
  show that the eccentricities of the growing terrestrial planets remain low
  throughout the simulation, validating our assumption of pebble accretion on
  circular orbits in the main paper.}
  \label{f:mass_sma}
\end{figure*}

The results are shown in Figure \ref{f:mass_sma}. The growth tracks largely
follow the single-planet growth tracks presented in the main text. There is only
little interaction between Venus, Earth and Theia while they grow, exciting
eccentricities up to $e \sim 10^{-2}$. The planets thus remain on fairly
circular orbits, which validates our assumption of pebble accretion on circular
orbits in the main text. Overall we conclude that the $N$-body simulations give
similar results to the single-planet growth tracks presented in the main text.

\newpage

\section{Supplementary Figures}

\begin{figure*}[!h]
  \begin{center}
    \includegraphics[width=0.9\linewidth]{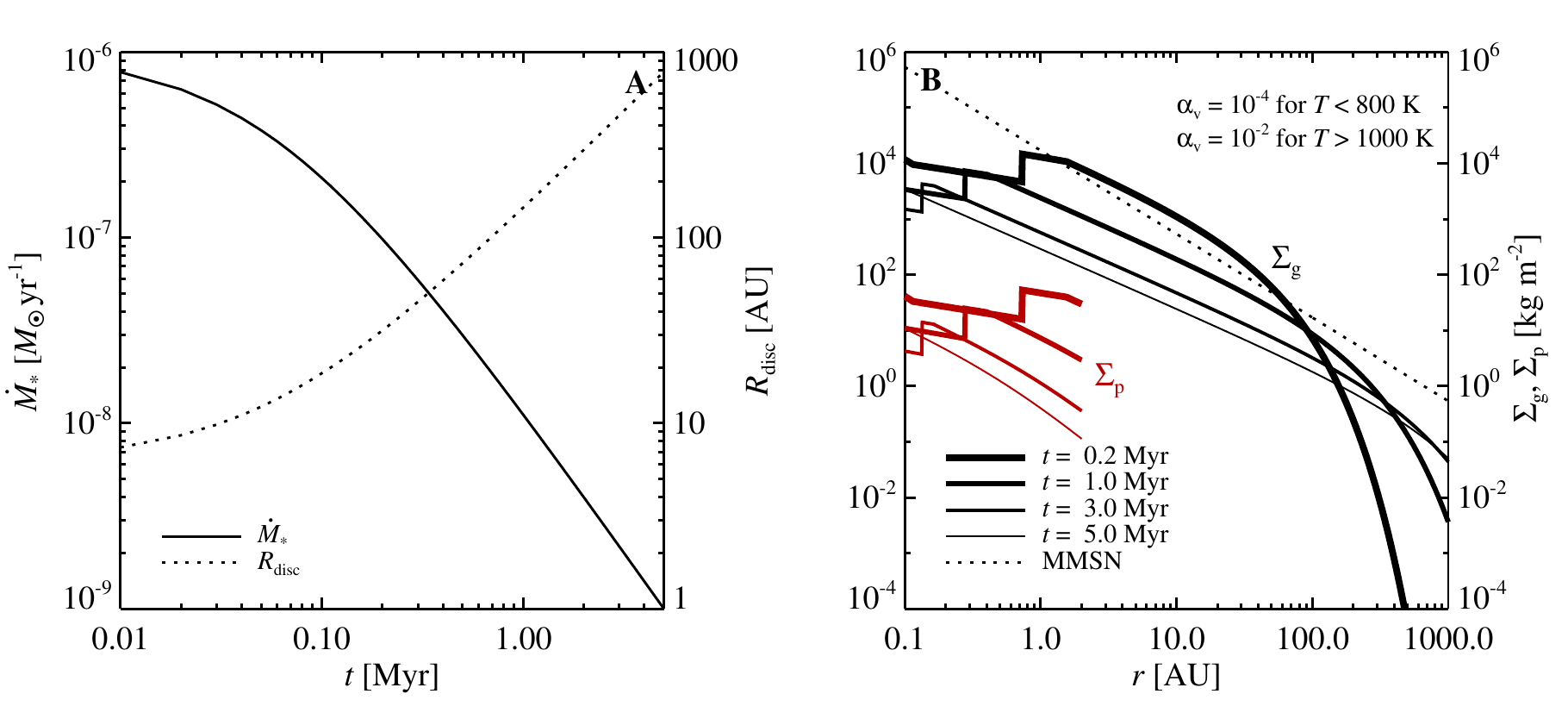}
  \end{center}
  \caption{{\bf Details of the protoplanetary disc model that includes
  heating by the magnetorotational instability above $T=800$ K.} Panel (A) shows
  the stellar accretion rate as a function of time (left axis) and the
  characteristic disc size (right axis). Panel (B) shows the column density
  profile of gas and pebbles at four different evolution times, as well as the
  minimum mass solar nebula for comparison (dotted line). The inner regions are
  viscously heated by the magnetorotational instability and hence the column
  density is lowered to maintain a constant accretion rate onto the star. We
  show the pebble column density only interior of 2 AU, since
  $\varSigma_{\rm p}$ is constructed for the terrestrial planet zone using a
  constant ratio ($\xi=0.0036$) of the pebble flux relative to the gas flux from
  the outer protoplanetary disc.}
  \label{f:disc_structure}
\end{figure*}

\newpage

\begin{figure*}[!h]
  \begin{center}
    \includegraphics[width=0.49\linewidth]{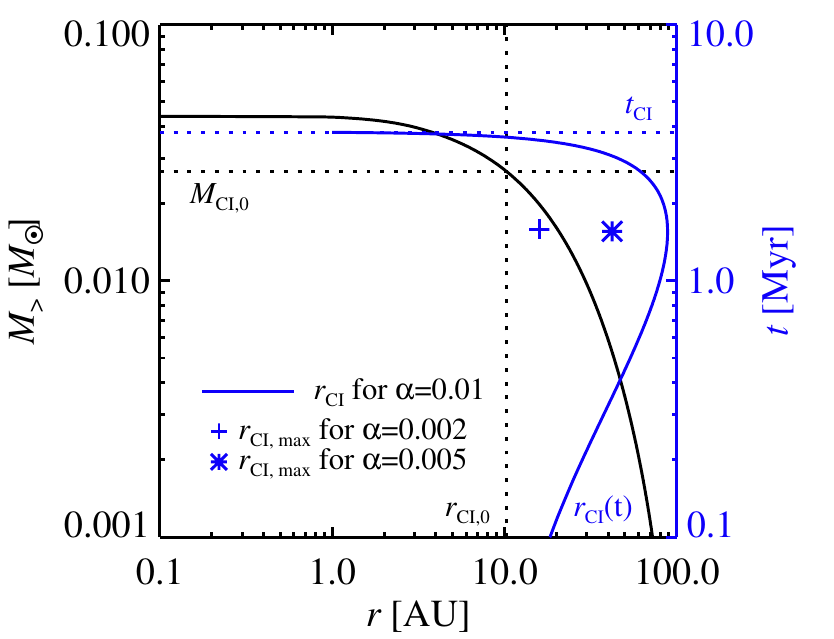}
  \end{center}
  \caption{{\bf The cumulative gas mass in the protoplanetary disc as a
  function of distance from the star} at a time of $t=10^5$ yr (left axis) and
  the location of the transition from NC to CI composition as a function of time
  (right axis). The CI material is required here to enter the terrestrial
  planet region at a time of $t=3.8\,{\rm Myr}$. This gives an initial
  transition line at approximately 10 AU at $t=0$ (which has already expanded to
  20 AU at $t=10^5$ yr where the plot starts). The initial mass of gas and dust
  in the CI region is 0.027 $M_\odot$ of the 0.044 $M_\odot$ total mass residing
  in the protoplanetary disc at $t=10^5\,{\rm yr}$. The plus and the asterisk
  symbols mark the maximum distance of the NC-CI interface curve for two lower
  values of $\alpha$: $16$ AU for $\alpha=0.002$ and $42$ AU for
  $\alpha=0.005$.}
  \label{f:alpha_disc_trajectories}
\end{figure*}

\newpage

\end{document}